\documentclass[useAMS,usenatbib]{mn2e}
\usepackage{graphicx}
\usepackage{amsfonts}
\usepackage{pifont}

\def\nyu{\mbox{NYU-VAGC}}

\usepackage{color} 
\definecolor{purple}{RGB}{160,32,240}
\usepackage {comment}
\usepackage{url}
\newcommand{\rockstar}{\textsc{Rockstar}}
\newcommand{\ctrees}{\textsc{Consistent Trees}}

\def\msun{\mbox{M$_{\odot}$}}

\def\gsmf{\mbox{GSMF}}
\def\phigal{\mbox{$\phi_{\rm gal}$}}

\def\Vmax{\mbox{$\mathcal{V}_{\rm max}$}}
\def\Vmaxj{\mbox{$\mathcal{V}_{{\rm max},i}$}}
\def\vmax{\mbox{$V_{\rm max}$}}
\def\vpeak{\mbox{$V_{\rm peak}$}}
\def\mvir{\mbox{$M_{\rm vir}$}}
\def\ms{\mbox{$M_*$}}
\def\ugriz{\mbox{$ugriz$}}

\defcitealias{RDA12}{RDA12}
\def\ltsima{$\; \buildrel < \over \sim \;$}    % Use in text mode
\def\lesssim{\lower.5ex\hbox{\ltsima}}           % Use in math mode
\def\gtsima{$\; \buildrel > \over \sim \;$}    % Use in text mode
\def\grtsim{\lower.5ex\hbox{\gtsima}}           % Use in math mode

\title[Does the Galaxy-Halo Connection Vary with Environment?]
 {Does the Galaxy-Halo Connection Vary with Environment?}
\author[]{Radu Dragomir$^{1}$\thanks{radragomir@gmail.com},  Aldo Rodr\'iguez-Puebla$^{2,3}$\thanks{apuebla@astro.unam.com}, Joel R. Primack$^1$\thanks{joel@ucsc.edu }, Christoph T. Lee$^{1}$\thanks{christoph28@gmail.com}
\\
$^1$Physics Department, University of California, Santa Cruz, CA 95064, USA \\
$^2$ Department of Astronomy \&\ Astrophysics, University of California at Santa Cruz, Santa Cruz, Ca 95064, USA\\
$^3$ Instituto de Astronom\'ia, Universidad Nacional Aut\'onoma de M\'exico, A. P. 70-264, 04510, M\'exico, D.F., M\'exico\\
}

\date{Released 20?? Xxxxx XX}

\pagerange{\pageref{firstpage}--\pageref{lastpage}} \pubyear{20??}

\begin{document}

\label{firstpage}

\maketitle

\begin{abstract}
SubHalo Abundance Matching (SHAM) assumes that one (sub)halo property, such as mass $\mvir$ or peak circular velocity $\vpeak$, determines properties of the galaxy hosted in each (sub)halo such as its luminosity or stellar mass. 
This assumption implies that the dependence of Galaxy Luminosity Functions (GLFs) and the Galaxy Stellar Mass Function (GSMF) on environmental density is determined by the corresponding halo density dependence. In this paper, we test this by determining from an SDSS sample the observed dependence with environmental density of the \ugriz\ GLFs and GSMF for all galaxies, and for central and satellite galaxies separately. We then show that the SHAM predictions are in remarkable agreement with these observations, even when the galaxy population is divided between central and satellite galaxies. However, we show that SHAM fails to reproduce the correct dependence between environmental density and $g-r$ color for all galaxies and central galaxies, although it better reproduces the color dependence on environmental density of satellite galaxies. 
\end{abstract}


\section{Introduction}

In the standard theory of galaxy formation in a $\Lambda$CDM universe, 
galaxies form and evolve in massive dark matter halos. 
The formation of dark matter halos is through two main mechanisms: (1) 
the accretion of diffuse material, and (2) the incorporation of material when halos merge. 
At the same time, galaxies evolve within these halos, where multiple physical mechanisms 
regulate star formation and thus produce their observed properties. Naturally, 
this scenario predicts that galaxy properties are influenced by the formation and evolution of their host
halos \citep[for a recent review see][]{Somerville+2015}. 

What halo properties matter for galaxy formation?  The simplest assumption that
galaxy formation models make is that a dark matter halo property such as mass $\mvir$ or maximum circular velocity $\vmax$ {\it fully determines the statistical properties}
of their host galaxies. This assumption was supported by early studies that showed that 
the halo properties strongly correlate with the larger-scale environment mainly due to changes
in halo mass \citep[e.g.,][]{Lemson+1999}. Halo evolution and corresponding evolution of galaxy properties can be predicted from Extended Press-Schechter analytical models based on Monte Carlo merger trees \citep{Cole1991,White+1991,Kauffmann+1993,Somerville+1999}.\footnote{More recent methods apply corrections that improve agreement with N-body simulations \citep{Parkinson+2008,Somerville+2015}.} Such models assume
that the galaxy assembly time and merger history are independent of the large-scale environment \citep[for 
a recent discussion see, e.g.,][]{Jiang+2014}.

However, it is known that dark matter halo properties do depend on other aspects beyond $\mvir$, a phenomenon known as halo assembly bias. \citet[][see also \citealp{Gao+2005,Gao+2007,Faltenbacher+2010,Lacerna+2011}]{Wechsler+2006} observed an assembly bias effect in the clustering of 
dark matter halos: they showed that for halos with $\mvir\lesssim10^{13}\msun$ early forming halos are more
clustered than late forming halos, while for more massive halos they found the opposite. Other effects of environmental
density on dark matter halos are known, for example that halo mass accretion rates and spin can be significantly reduced in dense environments due to tidal effects, and that median halo spin is significantly reduced in low-density regions due to the lack of tidal forces there \citep{Lee+2017}. 
%{\bf
Indeed there are some recent efforts to study 
assembly bias and the effect of the environment on the galaxy-halo connection in the context of galaxy clustering \citep{Lehmann+2017,VakiliHahn2016,Zentner+2016,Zehavi+2017} and weak lensing \citep{Zu+2017}.
%}
Despite such environmental effects on halo properties, it may still be true that some galaxy properties can be correctly predicted from just halo $\mvir$ or $\vmax$.

The assumption that dark matter halo mass fully determines the statistical properties
of the galaxies that they host has also influenced the development of empirical approaches 
for connecting galaxies to their host halo: the so-called halo occupation distribution (HOD)
models  \citep{Berlind+2002}  and the closely related conditional stellar mass/luminosity function model
%in which the number of galaxies is a function of stellar mass/luminosity 
\citep{Yang+2003,Cooray+2006}.
HOD models assume that the distribution of galaxies depends on halo mass {\it only} \citep{Mo+2004,Abbas+2006}. 
Yet the HOD assumption has 
been successfully applied to explain the clustering properties of galaxies not only as a function
of their mass/luminosity only but also as a function of galaxy colors \citep{Jing+1998,Berlind+2002,Zehavi+2005,Zheng+2005,Tinker+2013,Rodriguez-Puebla+2015}. 

The (sub)halo abundance 
matching (SHAM) approach takes the above assumption to the next level by assuming that not only does a halo property, such as mass $\mvir$ or maximum circular velocity $\vmax$, determine the luminosity or stellar mass of central galaxies, but also that {\it there is a simple relation between subhalo properties and those of the satellite galaxies they host}. Specifically, we will assume that 
subhalo peak circular velocity $\vpeak$ fully determines the corresponding properties of their hosted satellite galaxies \citep{Reddick+2013}. For simplicity, in the remainder of this paper, when we write $\vmax$ we will mean the maximum circular velocity for distinct halos, and the peak circular velocity of subhalos.
SHAM assigns by rank a halo property, such as $\vmax$, 
to that of a galaxy property, such as luminosity or stellar mass, by matching their 
corresponding cumulative number densities 
\citep{Kravtsov+2004,ValeOstriker2004,Conroy+2006,Conroy+2009,Behroozi+2010,Behroozi+2013a,Moster+2013,Moster+2017,RP17}. 

%In general, the galaxy population can be divided into two categories; centrals and satellite galaxies. 
While central galaxies are continuously growing by in-situ star formation and/or galaxy mergers,
satellite galaxies are subject to {\it environmental effects} such as tidal and ram-pressure stripping,
in addition to interactions with other galaxies in the halo and with the 
halo itself. Therefore, central and satellite galaxies are expected to differ in the
relationship between their host halos and subhalos \citep[see e.g.,][]{Neistein+2011,RDA12,Yang+2012,RAD13}. 
Nevertheless, SHAM assumes that (sub)halo $\vmax$ fully determines the statistical properties of the galaxies.
Thus SHAM galaxy properties evolve identically for central and satellite galaxies, except that satellite galaxy properties are fixed after $\vpeak$ is reached.\footnote{Note that subhalo $\vpeak$ is typically reached not at accretion, but rather when the distance of the progenitor halo from its eventual host halo is 3-4 times the host halo $R_{\rm vir}$ \citep{Behroozi14}.}   
SHAM also implies that galaxy properties are independent of local as well as large-scale environmental densities. Thus two halos
with identical \vmax\ but in different environments will host identical galaxies. Despite the 
extreme simplicity of this approach, the two point correlation functions predicted by SHAM are in excellent
agreement with observations \citep[][and Figures \ref{wrp_lum} and \ref{wrp_ms} below]{Reddick+2013,Campbell+2017}, showing that {\it on average} galaxy clustering depends on halo $\vpeak$. It is worth mentioning that 
neither HOD nor SHAM identify clearly which galaxy property, luminosity in various 
wavebands or stellar mass, depends more strongly on halo mass---although, theoretically, 
stellar mass growth is expected to be more closely related to halo mass accretion \citep{SHARC}. 

Our main goal in this paper is to determine whether the assumption that one (sub)halo property, in our case
halo $\vmax$ and subhalo $\vpeak$, fully determines some statistical properties of the hosted galaxies. This might be true even though the galaxy-halo relation is expected to depend on environment because the properties of the galaxies might reflect halo properties that depend on some environmental factor \citep[see e.g.,][]{Lee+2017}. We will test this assumption by determining from a Sloan Digital Sky Survey (SDSS) sample the dependence on environmental density of the \ugriz\ galaxy luminosity functions (GLFs) as well as the Galaxy Stellar Mass Function (\gsmf) for all galaxies, and separately for central and satellite galaxies, and comparing these observational results with SHAM predictions. We will also investigate which of these galaxy properties is better predicted by SHAM.  
If a galaxy-halo connection that is independent of environment successfully reproduces observations in the nearby universe, then we can conclude that the relation may be appropriate to use for acquiring other information about galaxies. It also suggests that this assumption be tested at larger redshifts. 
To the extent that the galaxy-halo connection is independent of density or other environmental factors, it is a great simplification. 
%A universal relation would also imply that there is a single halo property that fully determines galaxy formation.

This paper is organized as follows. In Section \ref{Obs_data}, we describe the galaxy sample that we utilize for the determination 
of the environmental dependence of the \ugriz\ GLFs and \gsmf. Section \ref{SHAM} describes our mock galaxy
catalog based on the Bolshoi-Planck cosmological simulation. Here we show how SHAM assigns to every halo in the simulation has five band
magnitudes, \ugriz, and a stellar mass. In Section \ref{results}, we present the dependence with environment of  
 \ugriz\ GLFs and \gsmf\ both for observations and for SHAM applied to the Bolshoi-Planck simulation. We show that the SHAM predictions are in remarkable agreement with observations even when the galaxy population is divided between central and satellite galaxies. However, we also find that SHAM fails to reproduce the correct dependence between environmental density and $g-r$ color. Finally Section \ref{summ_discussion} summarizes our results and discusses our findings. We adopt a \citet{Chabrier2003} IMF and the Planck cosmological parameters used in the Bolshoi-Planck simulation: $\Omega_\Lambda=0.693,\Omega_{\rm M}=0.307, h=0.678$.

\section{Observational Data}
\label{Obs_data}

In this section we describe the galaxy sample that we utilize for the determination of the galaxy distribution. 
We use the standard 1/\Vmax\ weighting procedure for the determination of the \ugriz\ Galaxy Luminosity Functions (GLFs) and the Galaxy Stellar Mass Function (\gsmf) and report their corresponding best fitting models. 
We show that a function composed of a single Schechter function plus another Schechter function with a 
sub-exponential decreasing slope is an accurate model for the \ugriz\ GLFs as well as the \gsmf. 
Finally, we describe the methodology for the determination of the environmental density
dependence of the \ugriz\ GLFs and \gsmf. 

\subsection{The Sample of Galaxies}

In this paper we utilize the New York Value Added Galaxy Catalog \nyu\ \citep{Blanton+2005a} based on the 
the SDSS DR7. Specifically, we use the large galaxy group catalog from \citet{Yang+2012}\footnote{This galaxy group 
catalog represents an updated version of \citet{Yang+2007}; see also \citet{Yang+2009b}.} with $\sim6\times10^{5}$ 
spectroscopic galaxies over a solid angle of $7748$ deg$^2$ comprising the redshift range 
$0.01 < z < 0.2$ with an apparent magnitude limit of $m_{{\rm lim}, r} = 17.77$. 
However, the sample we use in this paper is $0.03 < z < 0.11$ (see Figure \ref{Mag_vs_z} below).

The \citet{Yang+2012} catalog is a large halo-based galaxy group catalog that assigns group membership 
by assuming that the distribution of galaxies in phase space follows that of dark matter particles. 
Mock galaxy catalogs demonstrate that $\sim80\%$ of all their groups have a completeness larger 
than $80\%$ while halo groups with mass $M_{\rm vir}>10^{12.5}h^{-1}\msun$ have a completeness 
$>95\%$; for more details see \citet{Yang+2007}. 
Here, we define central galaxies as the most massive galaxy in their group in terms 
of stellar mass; the remaining galaxies will be regarded as satellites. 

The definition of groups in the \citet{Yang+2012} catalog is very broad and includes 
systems that are often explored individually in the literature, such as clusters, compact groups, 
fossil groups, rich groups, etc. That is, this galaxy group catalog is not biased 
to a specific type of group. Instead, this galaxy group catalog is diverse and, more importantly, closely related 
to the general idea of galaxy group that naturally emerges in the $\Lambda$CDM paradigm: that halos host
a certain number of galaxies inside their virial radius. Therefore, the \citet{Yang+2012} galaxy group catalog 
is ideal for comparing to predictions based on $N$-body cosmological simulations. For
the purpose of exploring whether certain galaxy properties are fully determined by the (sub)halo in which they reside, this 
galaxy group catalog will help us to draw conclusions not only at the level of the global 
GLFs and \gsmf\ but also at the level of centrals and satellites. Thus, the \citet{Yang+2012} galaxy group catalog
is an ideal tool to explore at a deeper level the simple assumptions in the SHAM approach.  

%%%%%%%%%%%%%%%%%%%%%%%%%%%%%%%
\begin{figure*}
\includegraphics[height=3.8in,width=5.2in]{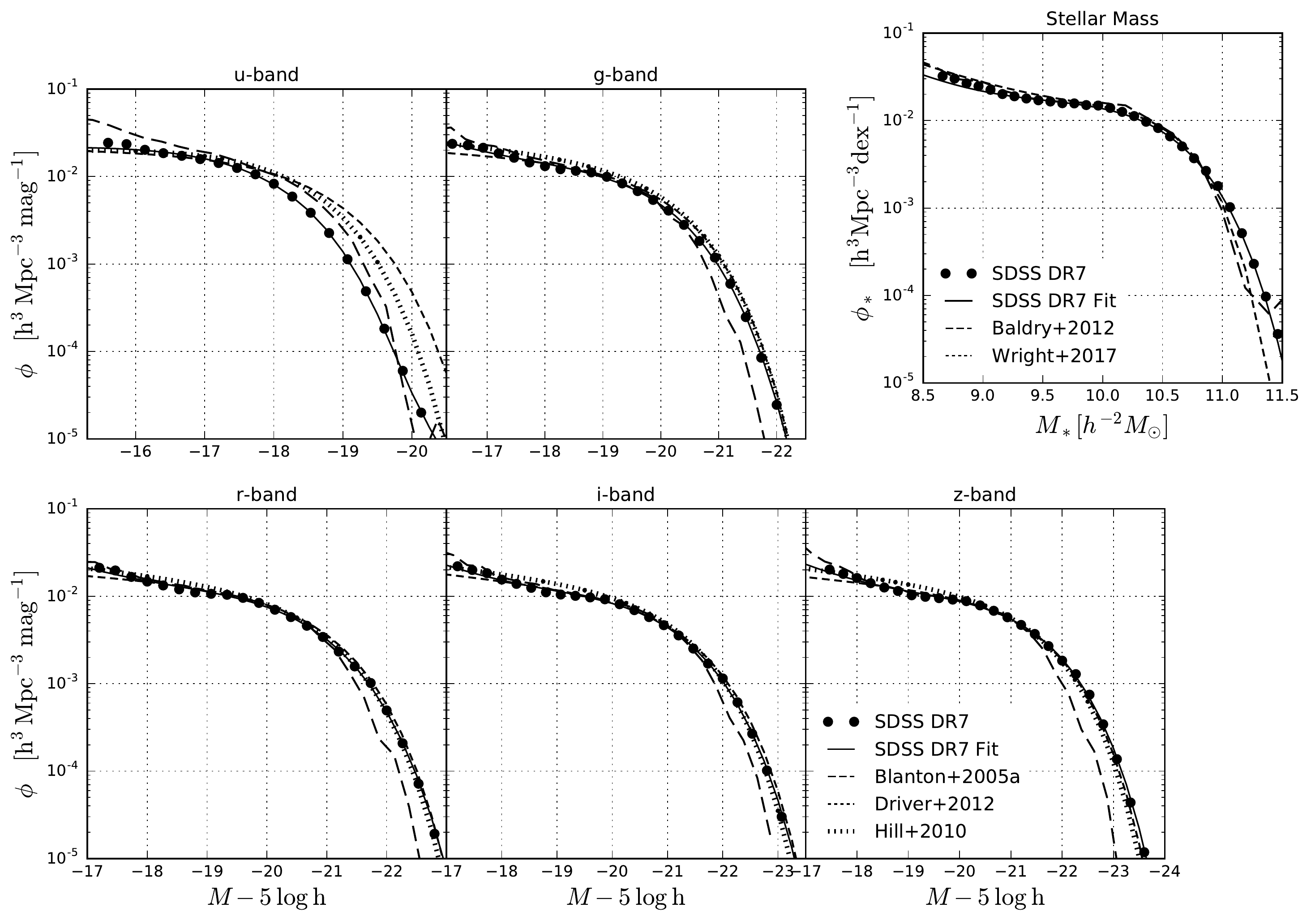}
\caption{ The global \ugriz\ galaxy luminosity function. Our derived \ugriz\ GLFs and \gsmf\ are shown with the black circles with error bars. 
For comparison we reproduce the \ugriz\ GLFs from \citet[][black long dashed lines]{Blanton+2005} based on the SDSS DR2; \citet[][dotted lines]{Hill+2010}
by combining the MGC, SDSS DR5 and the UKIDSS surveys; and \citet[][short dashed lines]{Driver+2012} based on the GAMA survey.
As for the stellar masses we compare with the \gsmf\ from \citet{Baldry+2012} and \citet{Wright+2017}, black long and short dashed lines, respectively.
 }
\label{global_GLF}
\end{figure*}
%%%%%%%%%%%%%%%%%%%%%%%%%%%%%%%

 %%%%%%%%%%%%TABLE%%%%%%%%%%%%%%%%%%%%%%
\begin{table*}
	\caption{Best fitting parameters for the GLFs and the GSMF.}
	\begin{center}
		\begin{tabular}{c c c c c c}
			\hline
						\multicolumn{5}{c}{Galaxy Luminosity Functions} \\  
			\hline
			\hline
			Band & $\alpha$ & $M^*-5\log h$ & $\log\phi^*_1$ $\left[h^3 {\rm Mpc}^{-3} {\rm mag}^{-1}\right]$ & $\log\phi_2^*$  $\left[h^3 {\rm Mpc}^{-3} {\rm mag}^{-1}\right]$ & $\beta$ \\
			\hline
			\hline
			$u$ & $-0.939 \pm 0.005$ & $-17.758\pm0.016$ & $-1.530 \pm 0.002$ & $-3.692 \pm 0.044$  & $0.721 \pm 0.008$\\
			$g$ & $-1.797 \pm 0.044$ & $-19.407\pm0.068$ & $-2.764 \pm 0.105$ & $-1.674 \pm 0.013$ & $0.821 \pm 0.014$\\
			$r$ & $-1.810 \pm 0.036$ &  $-20.184\pm0.062$ & $-2.889 \pm 0.094$ & $-1.733 \pm 0.013$ & $0.813 \pm 0.013$\\
			$i$ & $-1.794 \pm 0.031$ & $-20.546\pm0.053$ & $-2.896 \pm 0.077$ & $-1.768 \pm 0.011$ & $0.815 \pm 0.011$\\
			$z$ & $-1.816 \pm 0.028$ & $-20.962\pm0.051$ & $-3.038 \pm 0.076$ & $-1.806 \pm 0.012$ & $0.827 \pm 0.011$\\
			\hline
									\multicolumn{5}{c}{Galaxy Stellar Mass Function} \\  
			\hline
			\hline
			 & $\alpha$ & $\mathcal{M}^*$ $\left[h^{-2} \msun\right]$ & $\log\phi^*_1$ $\left[h^3 {\rm Mpc}^{-3} {\rm dex}^{-1}\right]$ & $\log\phi_2^*$  $\left[h^3 {\rm Mpc}^{-3} {\rm dex}^{-1}\right]$ & $\beta$ \\
			\hline
			 & $-1.664 \pm 0.033$ & $10.199\pm0.0303$ & $-3.041 \pm 0.082$ & $-1.885 \pm 0.010$ & $0.708 \pm 0.012$ \\
			\hline
			
		\end{tabular}
		\end{center}
	\label{T1}
\end{table*}
%%%%%%%%%%%%%%%%%%%%%%%%%%%%%%%%%%%%%

In order to allow for meaningful comparison between galaxies at different redshifts, we utilize model 
magnitudes\footnote{Note that we are using model magnitudes instead of Petrosian magnitudes. The main 
reason is because the former ones tend to underestimate the true light from galaxies particularly for high-mass 
galaxies, see e.g., \citet{Bernardi+2010} and \citet{Montero-Dorta+2009}\label{foot1}.} that are 
K+E-corrected at the rest-frame $z=0$. These corrections account for the broad band shift with respect to 
the rest-frame broad band and for the luminosity evolution. For the K-corrections we utilize the input 
values tabulated in the \nyu\ catalog \citep[][corresponding to the  
\textsc{kcorrect} software version v$4\_1\_4$]{Blanton+2007b}, while for the evolution term we assume a 
model given by
\begin{equation}
E_j(z) = -Q_X \times z,
\label{Ecorr}
\end{equation}
where the subscript $X$ refers to the $u$, $g$, $r$, $i$ and $z$ bands and their values are 
$(Q_u, Q_g, Q_r, Q_i, Q_z) = (4.22, 1.3, 1.1, 1.09, 0.76)$. Here we ignore potential dependences between
$Q_X$ and colors \citep[but see][ for a discussion]{Loveday+2012} and luminosity, and use global values only.  
Although this is a crude approximation for accounting for the evolution of the galaxies, it is accurate
enough for our purposes since we are not dividing the galaxy distribution into subpopulations as a function of 
star formation rate and/or color. 

We estimated the value of each $Q_X$ by determining 
first the $X$-band GLF when $Q_X=0$ at four redshift intervals: $[0.01,0.05]$,  $[0.01,0.1]$, $[0.01,0.15]$ and 
$[0.01,0.2]$. When assuming $Q_X=0$, the GLFs are normally shifted towards higher luminosities, with this shift
increasing with redshift. In other words, when ignoring the evolution correction, the GLF will result in an
overestimation of the number density at higher luminosities and high redshifts. Thus, in order to account for 
this shift we find the best value for $Q_X$ that leaves the GLFs invariant at 
the four redshift intervals mentioned above. We note that our derived values are similar
to those reported in \citet{Blanton+2003b}. For the $u$-band we used the value reported in 
\citet{Blanton+2003b}, but we have checked that the value of $Q_u = 4.22$ also leaves the GLF 
invariant at the four redshifts bins mentioned above. 

For stellar masses, we utilize the MPA-JHU DR7 database derived from photometry-spectral 
energy distribution fittings, explained in detail in \citet{Kauffmann+2003}. All stellar masses have been 
normalized to a \citet{Chabrier2003} IMF and to the cosmology used for this paper. 

\subsection{The Global \ugriz\ Luminosity Functions and Stellar Mass Function}
\label{global_density}

Next, we describe the procedure we utilize for determining the global GLFs and the \gsmf. 

Here, we choose the standard 1/\Vmax\ weighting procedure for the determination of the \ugriz\ GLFs 
and the \gsmf. Specifically, we determine the galaxy luminosity and stellar mass distributions as
\begin{equation}
\phi_X(M_X) = \frac{1}{\Delta M_X}\sum_{i=1}^N\frac{\omega_X(M_X\pm \Delta M_X / 2)}{\Vmaxj},
\end{equation}
where $M_X$ refers to $M_u$, $ M_g$, $ M_r$, $ M_i$, $ M_z$ and $\log\ms$, $\omega_i$ is 
the correction weight completeness factor in the \nyu\ for galaxies within the interval 
$M_X\pm \Delta M_X / 2$, and 
\begin{equation}
\Vmaxj = \int_{\Omega}\int^{z_u}_{z_l}\frac{d^2V_c}{dzd\Omega} dzd\Omega. 
\end{equation}
We denote the solid angle of the SDSS DR7 with $\Omega$ while $V_c$ refers to the comoving 
volume enclosed within the redshift interval $[z_l, z_u]$. The redshift limits are defined as 
$z_l = {\rm max}(0.01, z_{\rm min})$ and $z_u = {\rm min}(z_{\rm max}, 0.2)$; where $z_{\rm min}$ 
and $z_{\rm max}$ are, respectively, the minimum and maximum at which each galaxy can be 
observed in the SDSS DR7 sample. For the completeness limits, we use the limiting
apparent magnitudes in the $r$-band of $r=14$ and $r=17.77$. 

The filled black circles with error bars in Figure \ref{global_GLF} present our determination of the global SDSS DR7 \ugriz\ GLFs. For comparison we reproduce the \ugriz\ GLFs from 
\citet[][black long dashed line]{Blanton+2005} who used a sample of 
low-redshift galaxies ($<150 h^{-1}$Mpc) from the SDSS DR2 and corrected due to low surface 
brightness selection effects. Additionally, we compare to \citet{Hill+2010} who combined data from
the Millennium Galaxy Catalog (MGC), the SDSS DR5 and the UKIRT Infrared Deep Sky Survey 
Large Area Survey (UKIDSS) for galaxies with $z<0.1$, dotted lines; and to \citet{Driver+2012} who 
utilized the Galaxy And Mass Assembly (GAMA) survey for the redshift interval $0.013<z<0.1$ to derive the \ugriz\ 
GLFs, short dashed-lines. All the GLFs in Figure \ref{global_GLF} are at the rest-frame $z=0$. 
In general we observe good agreement with previous studies; in a more detailed examination, however, 
we note some differences that are worthwhile to clarify. 

Consider the $u$-band GLFs from Figure \ref{global_GLF} and note that there is an 
apparent tension with previous studies.
At the high luminosity-end, our inferred $u$-band GLF decreases much faster than the above-mentioned
studies. This is especially true when comparing with the \citet{Hill+2010} and \citet{Driver+2012} GLFs. 
This could be partly due to the differences between the Kron magnitudes
used by \citealp{Hill+2010} and \citealp{Driver+2012} and the model 
magnitudes used in this paper. But we believe that most of the difference is due to the differences 
in the E-corrections, reflecting that our model evolution is more extreme 
than that of \citet{Hill+2010} and \citet{Driver+2012}. This can be easily understood 
by noting that the high luminosity-end
of the GLF is very sensitive to E-corrections. The reason is that brighter galaxies are expected
to be observed more often at larger redshifts than fainter galaxies; thus Equation (\ref{Ecorr}) will result in a 
small correction for lower luminosity galaxies (low redshift) but a larger correction for higher luminosity 
galaxies (high redshifts). Indeed, \citet{Driver+2012} 
who did not determine corrections by evolution, derived a $u$-band GLF that predicts the largest abundance 
of high luminosity galaxies. On the other hand, the evolution model introduced by \citet{Hill+2010} is shallower 
than ours, which results in a GLF between our determination and the \citet{Driver+2012} $u$-band
GLF. This could explain the apparent tension between the different studies. While the effects of 
evolution are significant in the $u$-band, they are smaller in the longer wavebands.
Ideally, estimates of the evolution should be more physically motivated by galaxy formation models, 
but empirical measurements are more accessible and faster to determine; however, when making comparisons
one should keep in mind that empirical estimates are by no means definitive. 

Some previous studies have concluded that a single Schechter function is consistent with 
observations \citep[see, e.g.,][and recently \citealp{Driver+2012}]{Blanton+2003b}. 
However, other studies have found that a double Schechter function
is a more accurate description of the GLFs  \citep{Blanton+2005}. Additionally, recent studies 
have found shallower slopes at the high luminosity-end instead of an exponential decreasing slope in the 
GLFs\footnote{Note that this is not due to sky subtraction issues, as previous studies have 
found \citep[see, e.g.,][]{Bernardi+2013,Bernardi+2016}, since we are not including this correction 
in the galaxy magnitudes.  Instead, it is most likely due to our use of model magnitudes instead of Petrosian 
ones, see also footnote \ref{foot1}.} \citep[see e.g.,][]{Bernardi+2010}. In this paper, we choose to use 
GLFs that are described by a function composed of a single Schechter function plus another Schechter function 
with a subexponential decreasing slope for the \ugriz\ bands given by
	\begin{eqnarray}
		\phi(M)=\frac{\ln 10}{2.5}\phi_1^*10^{0.4(M^*_1-M)\left(1+\alpha_1\right)}\exp\left(-10^{0.4\left(M^*_1-M\right)}\right)& &  \nonumber \\
+\frac{\ln 10}{2.5}\phi_2^* 10^{0.4(M^*_2-M)\left(1+\alpha_2\right)}\exp\left(-10^{0.4\left(M^*_2-M\right)\beta}\right).
	\end{eqnarray}
The units of the GLFs are $h^3$ Mpc$^{-3}$ mag$^{-1}$ while the input magnitudes have units of mag$-5\log h$.  The parameters for the \ugriz\ bands are given in Table \ref{T1}. 
Note that for simplicity we assume that $\alpha_1 = \alpha $, $\alpha_2 = 1 + \alpha$ and 
$M^*_1 = M^*_2  =  M^* $.  These assumptions reduce the number of free parameters to five. 
The corresponding best fitting 
models are shown in Figure \ref{global_GLF} with the solid black lines. 
The filled circles with error bars in  
Figure \ref{global_GLF} present our determinations for the global SDSS DR7 GLFs. 

In the case of the \gsmf, we compare our results with \citet{Baldry+2012} and \citet{Wright+2017} plotted 
with the black long and short dashed lines respectively. Both analyses used the GAMA survey to determine 
the local \gsmf. Recall that our stellar masses were obtained from the MPA-JHU DR7 database. As can be seen in the figure, our determination is consistent with these previous results. 
We again choose to use a function composed of a single Schechter function plus another 
Schechter function with a subexponential decreasing slope for the \gsmf given by
\begin{eqnarray}
	\phi_*(\ms) =\phi_1^*{\ln 10}\left(\frac{\ms}{\mathcal{M}_1^*}\right)^{1+\alpha_1}\exp\left(-\frac{\ms}{\mathcal{M}_1^*}\right)& &  \nonumber \\
		+\phi_2^*{\ln 10}\left(\frac{\ms}{\mathcal{M}_2^*}\right)^{1+\alpha_2}\exp\left[-\left(\frac{\ms}{\mathcal{M}_2^*}\right)^\beta\right].
\label{mod_schec}
\end{eqnarray}
The units for the \gsmf\ are $h^3$ Mpc$^{-3}$ dex$^{-1}$ while the input stellar masses are in units of
$h^{-2}\msun$. Again, for simplicity we assume that  $\alpha_1 = \alpha$, $\alpha_2 = 1 + \alpha$, and
$\mathcal{M}_1^* = \mathcal{M}_2^* = \mathcal{M}^*$; again, this assumption reduces the number of free 
parameters to five. We report the best fitting value parameters in Table \ref{T1} and the corresponding 
best fitting model is presented with the solid black line in Figure \ref{global_GLF}. 
As we will describe in Section \ref{SHAM}, we use the \ugriz\ GLFs and GSMF as  
inputs for our mock galaxy catalog. 

\subsection{Measurements of the Observed \ugriz\ GLFs and GSMF as
a Function of Environment}

Once we determined the global \ugriz\ GLFs and the \gsmf, the next step in our program is to 
determine the observed dependence of the \ugriz\ GLFs and \gsmf\ with environmental density. 

\subsubsection{Density-Defining Population}

\begin{figure}
\hspace*{10pt}
\includegraphics[height=3in,width=3.in]{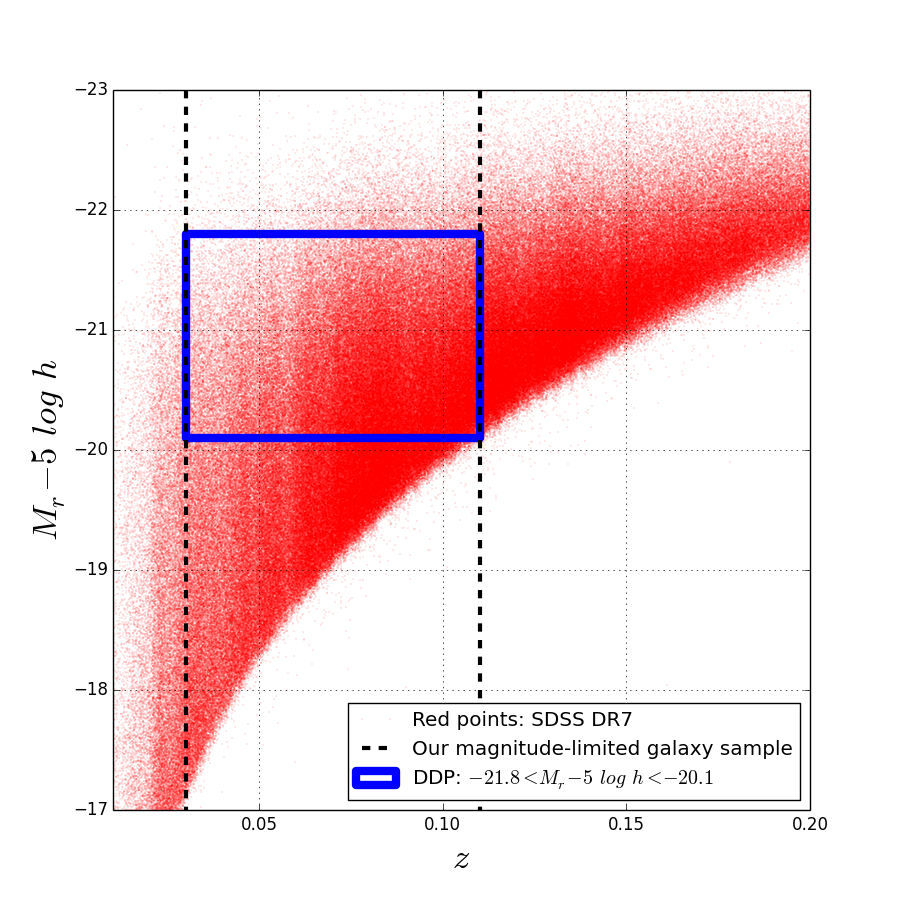}
\caption{Absolute magnitude in the $r-$band as a function of redshift for our 
magnitude-limited galaxy sample. The blue solid box shows our volume-limited DDP sample.
Note that our DDP sample restricts to study environments for galaxies between 
$0.03\leq z \leq0.11$ as shown by the dashed lines.
 }
\label{Mag_vs_z}
\end{figure}

The SDSS DR7 limiting magnitude in the $r$-band is 17.77. Thus, in order to determine the local 
overdensity of each SDSS DR7 galaxy, we need to first construct a volume-limited 
{\it density-defining population} (DDP, \citealp{Croton+2005, Baldry+2006}). A volume-limited sample 
can be constructed by defining the minimum and maximum redshifts at which galaxies within some
interval magnitude are detected in the survey. Following the \citet{McNaught-Roberts+2014} GAMA paper, 
we define 
our volume-limited DDP sample of galaxies in the absolute magnitude range $-21.8<M_r-5\log h<-20.1$.  
A valid question is whether the definition utilized for the volume-limited DDP sample
could lead to different results. This question has been studied in \citet{McNaught-Roberts+2014}; the authors 
conclude that the precise definition for the volume-limited DDP sample does not significantly affect the 
shape of GLFs. Nonetheless, our defined volume-limited DDP sample restricts the SDSS
magnitude-limited survey into the redshift range $0.03\leq z \leq0.11$. Figure \ref{Mag_vs_z} shows the 
absolute magnitude in the $r-$band as a function of redshift for our 
magnitude-limited galaxy sample. The solid box presents the galaxy population enclosed in our 
volume-limited DDP sample, while the dashed lines show our magnitude-limited survey. 

%%%%%%%%%%%%%%%%%%%%%%%%%%%%%%%
%%%%%%%%%%%%%%%%%%%%%%%%%%%%%%%

\subsubsection{Projected Distribution on the Sky of the Galaxy Sample}

The irregular limits of the projected distribution on the sky of the SDSS-DR7 galaxies 
could lead to a potential bias in our overdensity measurements; they will artificially increase the frequency of 
low density regions and, ideally, overdensity measurements should be carried out over more continuous regions. 
Following \citet{Varela+2012} and \citet{Cebrian_Trujillo2014}, we reduce this source of 
potential bias by restricting our galaxy sample to a projected area based on the following cuts:
\begin{equation}
	 {\rm DEC} > \left\{ 
			\begin{array}{l l}
				0 & \mbox{Southern limit}\\
				-2.555556\times({\rm RA} - 131^{\circ}) & \mbox{Western limit}\\
				-1.70909\times({\rm RA} - 235^{\circ}) & \mbox{Eastern limit}\\
				\arcsin\left(\frac{x}{\sqrt{1- x^2}}\right) & \mbox{Northern limit}
			\end{array},\right.
			\label{area_projection}
\end{equation}
where $x = 0.93232\sin({\rm RA} - 95.9^{\circ})$.  This region is plotted in Figure 1 of \citet{Cebrian_Trujillo2014}.
%{\color{red} Do we need a figure for this?}

\subsubsection{Overdensity Measurements}
\label{obs_Over}

In summary, our final magnitude-limited galaxy sample consists of galaxies in the redshift range 
$0.03\leq z\leq0.11$ and galaxies within the projected area given by Equation (\ref{area_projection}), while  
our volume-limited DDP sample comprises galaxies with 
absolute magnitude satisfying $-21.8<M_r-5\log h<-20.1$.
Based on the above specifications,
we are now in a position to determine the local overdensity of each SDSS DR7 galaxy in our
magnitude-limited galaxy sample. 

Overdensities are estimated by counting the number of DDP galaxy neighbors, $N_n$, 
around our magnitude-limited galaxy sample in spheres of 
$r_8 = 8h^{-1}$ Mpc radius. While there exists various methods to measure galaxy 
environments, \citet{Muldrew+2012} showed that aperture-based methods are more
robust in identifying the dependence of halo mass on environment, in contrast to
nearest-neighbors-based methods that are largely independent of halo mass.
In addition, aperture-based methods are easier to interpret. 
For these reasons, the aperture-based method is ideal to probe galaxy environments when testing
the assumptions behind the SHAM approach.  

The local density is simply defined as
\begin{equation}
\rho_8 = \frac{N_n}{4/3\pi r_8^3}.
\end{equation}
We then compare the above number to the expected number density of DDP galaxies 
by using the global $r$-band luminosity function determined above in Section \ref{global_density}; 
$\bar{\rho} = 6.094 \times 10^{-3} h^3$ Mpc$^{-3}$. Finally, the local density contrast for each galaxy
is determined as
\begin{equation}
\delta_8 = \frac{\rho_8 - \bar{\rho}}{\bar{\rho}}.
\label{delta_8}
\end{equation}
The effect of changing the aperture radius has been discussed in \citet{Croton+2005}. 
While the authors noted that using smaller spheres 
tends to sample underdense regions differently, they found that their conclusions 
remain robust due to the change of apertures. Nevertheless, smaller-scale
spheres are more susceptible to be affected by redshift space distortions. Following 
\citet{Croton+2005}, we opt to use spheres of $r_8 = 8h^{-1}$ Mpc radius as the best 
probe of both underdense and overdense regions. Finally, note that our main goal
is to understand whether halo \vmax\ fully determines galaxy properties as predicted by SHAM,
not to study the physical causes for the observed galaxy distribution with environment.
Therefore, as long as we treat our mock galaxy sample, to be described in Section \ref{SHAM}, 
in the same way that we treat observations, understanding the impact of changing apertures 
in the observed galaxy distribution is beyond the scope of this paper.  

\subsubsection{Measurements of the Observed \ugriz\ GLFs and the \gsmf\ as a Function of 
Environmental Density}

Once the local density contrast for each galaxy in the SDSS DR7 is determined, we estimate the 
dependence of the \ugriz\ GLFs and the \gsmf\ with environmental density. 
 
 As in Section \ref{global_density}, we use the standard 1/\Vmax\ weighting procedure. Unfortunately, 
 the  1/\Vmax\ method does not provide the effective volume covered by the overdensity bin in which the 
 GLFs and the \gsmf\ have been estimated and, therefore, one needs to slightly modify the 1/\Vmax\ 
 estimator. In this subsection, we describe how we estimate the effective volume. 
 
 We determine the fraction of effective volume by counting the number of DDP galaxy neighbours in a 
 catalog of random points with the same solid angle and redshift distribution as our final magnitude-limited 
 sample. Observe that we utilize the real position of the DDP galaxy sample defined above. 
 We again utilized spheres of  $r_8 = 8h^{-1}$ Mpc radius and create a random catalog consisting of $N_r\sim2\times10^{6}$ of points. 
 The local density contrast for each random point
is determined as in Equation (\ref{delta_8}):
\begin{equation}
\delta_{r_8} = \frac{\rho_{r_8} - \bar{\rho}}{\bar{\rho}},
\end{equation}
 where $\rho_{r_8}$ is the local density around random points. We estimate the fraction of 
 effective volume by a given overdensity bin as
 \begin{equation}
f(\delta_8) = \frac{1}{N_r}\sum_{i=1}^{N_r} \Theta(\delta_{{r_8},i}).
\label{over_fraction8}
\end{equation}
Here, $\Theta$ is a function that selects random points in the overdensity range $\delta_{r_8} \pm \Delta \delta_{r_8}/2$, that is:
\begin{equation}
	 \Theta(\delta_{{r_8},i}) = \left\{ 
			\begin{array}{c c}
				1 & \mbox{if } \delta_{{r_8},i} \in [\delta_{r_8} - \Delta \delta_{r_8}/2, \delta_{r_8} + \Delta \delta_{r_8}/2)\\
				0 & \mbox{otherwise } 
			\end{array}.\right.
\end{equation}
Table \ref{Tdensity} lists the fraction of effective volume for the range of overdensities considered in this paper and calculated as described
above. We estimate errors by computing the standard deviation of the fraction of effective volume in sixteen redshift bins equally spaced. 
We note that the number of sampled points gives errors that are less than $\sim3\%$ and for most of the
bins less than $\sim1\%$, see last column of Table \ref{Tdensity}. Therefore, we ignore any potential source of error from our determination of the
fraction of effective volume into the \ugriz\ GLFs and the $\gsmf$s. 

Finally, we modify the 1/\Vmax\ weighting estimator
to account for the effective volume by the overdensity bin as
\begin{equation}
\phi_X(M_X|\delta_8) = \sum_{i=1}^N\frac{\omega_i(M_X\pm \Delta M_X / 2|\delta_{r_8} \pm \Delta \delta_{r_8}/2)}{f(\delta_8)\times\Delta M_X\times\Vmaxj},
\end{equation}
again, $M_X$ refers to $M_u$, $ M_g$, $ M_r$, $ M_i$, $ M_z$ and $\log\ms$. Here $\omega_i$ refers 
to the correction weight completeness factor for galaxies within the interval $M_X\pm \Delta M_X / 2$
given that their overdensity is in the range $\delta_{r_8} \pm \Delta \delta_{r_8}/2$.

%%%%%%%%%%%%TABLE%%%%%%%%%%%%%%%%%%%%%%
\begin{table}
	\caption{Fraction of effective volume covered by the overdensity bins considered for our analysis in the SDSS DR7. Also shown is the 
	fractional error due to the number of random points sampled.}
	\begin{center}
		\begin{tabular}{c c c c}
			\hline			
			$\delta_{{\rm min},8}$ & $\delta_{{\rm max},8}$ & $f(\delta_8) \pm \delta f(\delta_8) $ & $100 \%\times \delta f(\delta_8) / f(\delta_8)$ \\
			\hline
			\hline
			-1 & -0.75 & $0.1963 \pm 0.0014$ & 0.713 \\
			-0.75 & -0.55 & $0.1094 \pm 0.0010$ &  0.914 \\
			-0.55 & -0.40 & $0.0974 \pm 0.0009$ &  0.924\\
			-0.40 & 0.00 & $0.2156 \pm 0.0014$  & 0.650 \\
			0.00 & 0.70 & $0.1800 \pm 0.0012$  & 0.667\\
			0.70 & 1.60 & $0.1040 \pm 0.0009$  & 0.866\\
			1.60 & 2.90 & $0.0621 \pm 0.0007$  & 1.130\\
			2.90 & 4 & $0.0197 \pm 0.0004$  &  2.030\\
			4.00 & $\infty$ & $0.0153 \pm 0.0004$ & 2.614 \\
			\hline
		\end{tabular}
		\end{center}
	\label{Tdensity}
\end{table}
%%%%%%%%%%%%%%%%%%%%%%%%%%%%%%%%%%%%%

\section{The galaxy-halo connection}
\label{SHAM}
%%%%%%%%%%%%%%%%%%%%%%%%%%%%%%%
\begin{figure*}
\includegraphics[height=2.8in,width=3.2in]{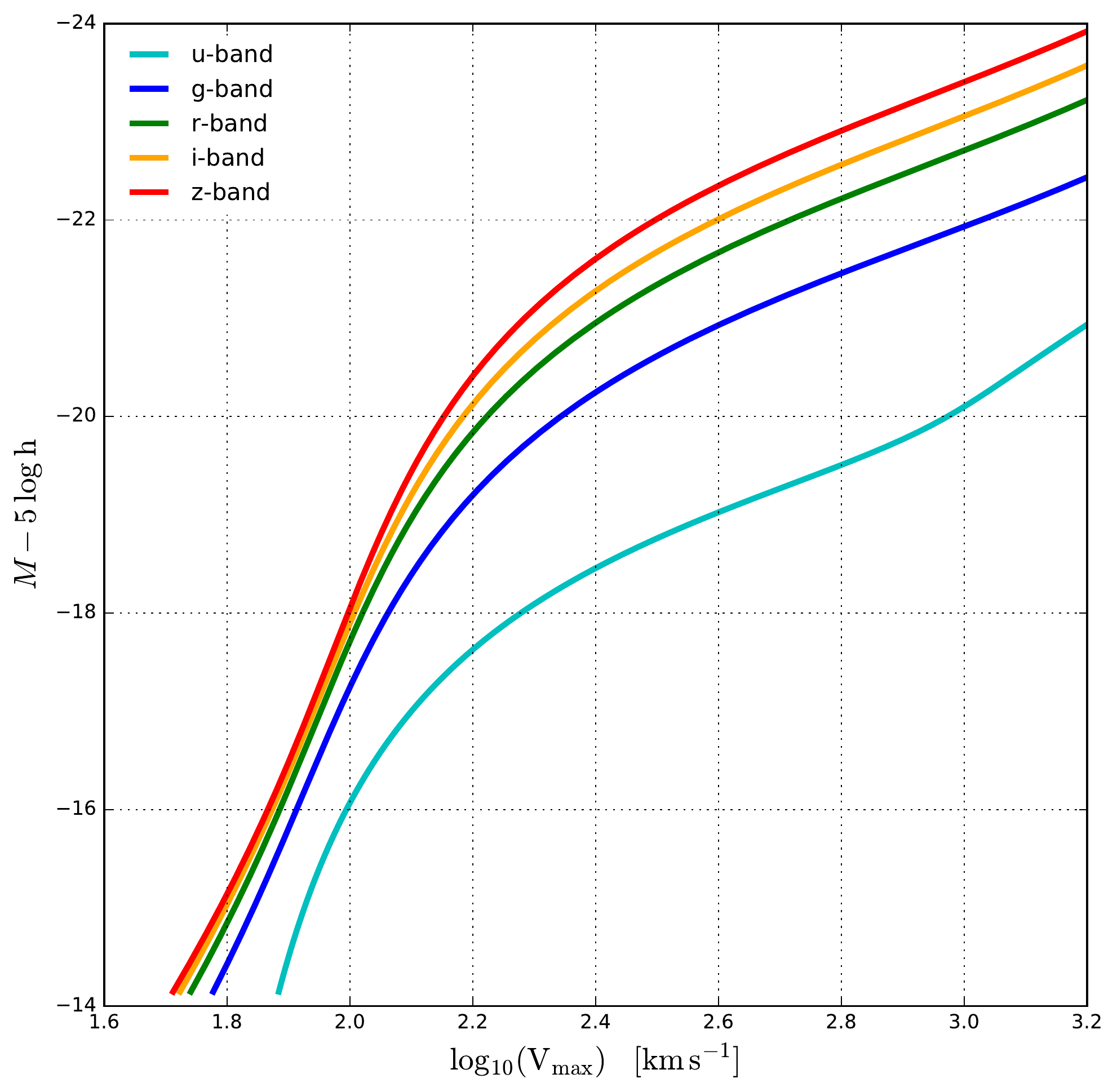}
\hspace*{10pt}
\includegraphics[height=2.8in,width=3.2in]{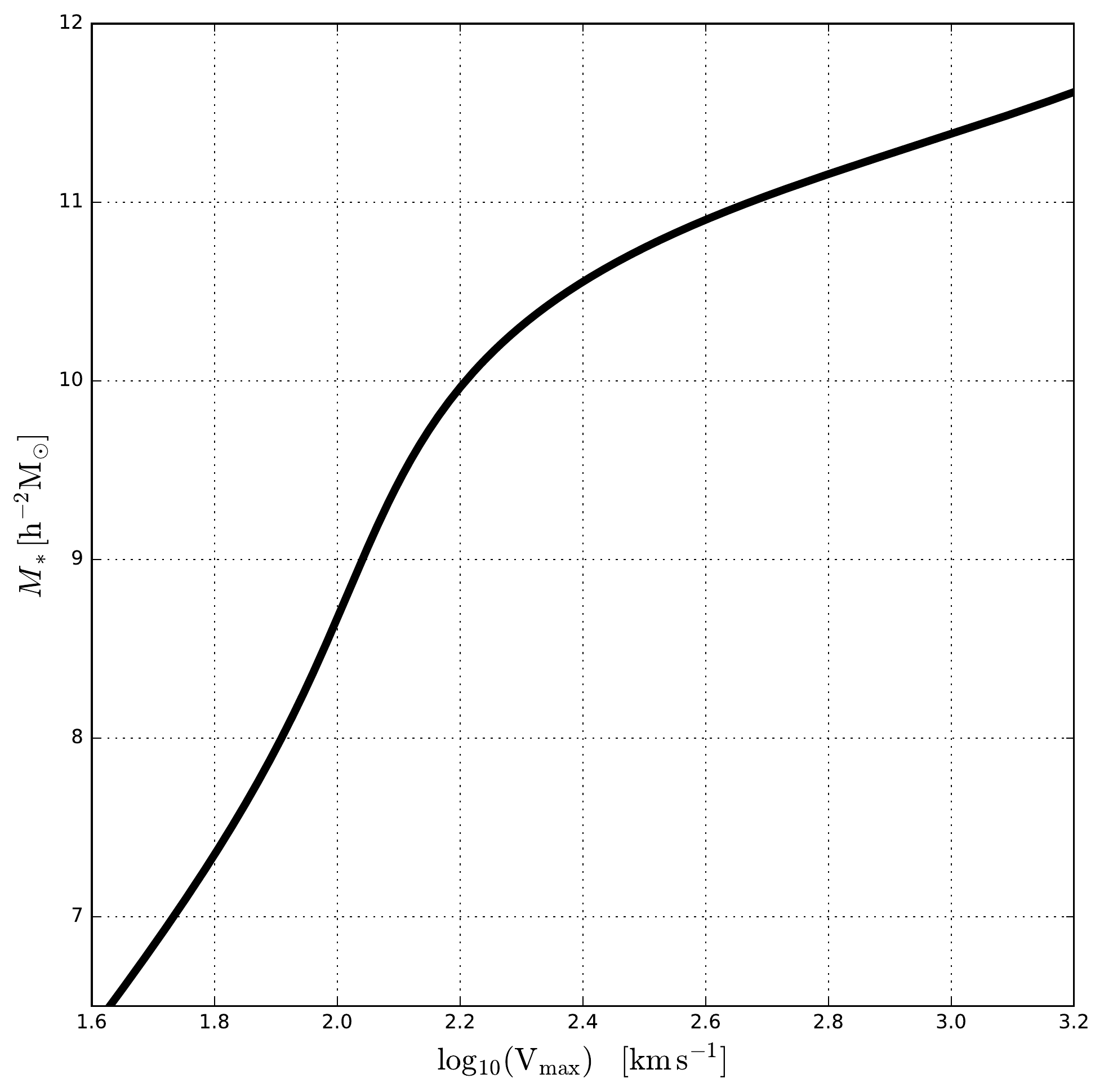}
\caption{{\bf Left Panel:} Luminosity-to-\vmax\ relation from SHAM. The different colors indicate the band utilized for the match. 
{\bf Right Panel:} Stellar mass-to-\vmax\ relation. Recall that SHAM assumes that these relations are valid for centrals as well 
as for satellites.  We report these values in Table \ref{galaxy_halo_table}.In the case of centrals \vmax\ refers to the halo maximum circular velocity, while for satellites $\vmax$ represents the highest maximum circular velocity ($\vpeak$)
reached along the subhalo's main progenitor branch. SHAM assumes that \vmax\ 
fully determines these statistical properties of the galaxies.}
\label{galaxy_halo_connection}
\end{figure*}
%%%%%%%%%%%%%%%%%%%%%%%%%%%%%%%

The main goal of this paper is to study whether one halo property, in this case \vmax, fully determines
%the statistical properties of the galaxies and thus 
the observed dependence with environmental density of the \ugriz\ GLFs and the \gsmf. 
Confirming this would significantly improve our understanding the galaxy-halo connection. 
In this section we describe 
%our determinations for the galaxy-halo connection and 
how we constructed a mock galaxy catalog in the cosmological  
Bolshoi-Planck N-body simulation via (sub)halo abundance matching (SHAM).

\subsection{The Bolshoi-Planck Simulation}
To study the environmental dependence of the galaxy distribution predicted by SHAM, we 
use the N-body Bolshoi-Planck (BolshoiP)  cosmological simulation \citep{Klypin+2016}. 
This simulation is based on the $\Lambda$CDM cosmology with parameters consistent 
with the latest results from the Planck Collaboration. This simulation has $2048^3$  particles 
of mass $1.9 \times 10^8 \msun h^{-1}$, in a box of side length $L_{\rm BP} = $ 250 $h^{-1}$Mpc. Halos/subhalos and their 
merger trees were calculated with the phase-space temporal halo finder \rockstar\ \citep{Behroozi+2013d} and the 
software \ctrees\ \citep{Behroozi+2013b}.
Entire \rockstar\ and  \ctrees\ outputs are downloadable.\footnote{\url{http://hipacc.ucsc.edu/Bolshoi/MergerTrees.html}} Halo masses 
were defined using spherical overdensities according to the redshift-dependent virial overdensity 
$\Delta_{\rm vir}(z)$ given by the spherical collapse model, with $\Delta_{\rm vir}(z) =333$ at $z=0$. 
The Bolshoi-Planck simulation is complete down to halos of maximum circular velocity $\vmax \grtsim 55$ km s$^{-1}$. For more
details see \citet{Rodriguez-Puebla+2016}. Next we describe our mock galaxy catalogs generated via SHAM.

\subsection{Determining the Galaxy-Halo Connection}
As we have explained, 
SHAM is a simple approach relating a halo property, such as mass or
maximum circular velocity, to that of a galaxy property, such as luminosity or stellar mass.  
%In its simplest form 
In abundance matching between a halo property and a galaxy property, the number density distribution of the 
halo property is matched to the number density distribution 
of the galaxy property to obtain the relation. Recall that SHAM
assumes that that there is a one-to-one monotonic relationship between galaxies and halos, and that centrals
and satellite galaxies have identical relationships (except that satellite galaxy evolution is stopped when the host halo reaches its peak maximum circular velocity).
In this paper we choose
to relate galaxy properties, $\mathcal{P}_{\rm gal}$, to halo maximum circular velocities \vmax\ as
\begin{equation}
\int_{\mathcal{P}_{\rm gal}}^{\infty} \phigal(\mathcal{P}_{\rm gal}') d\log\mathcal{P}_{\rm gal}' = \int_{V_{\rm max}}^{\infty} \phi_V (V_{\rm max} ') d\log V_{\rm max} ',
\label{AMT_Eq}
\end{equation}
where $\phigal(\mathcal{P}_{\rm gal})$ denotes the \ugriz\ GLF as well as the \gsmf\  and $\phi_V (\vmax)$ represents the
subhalo+halo velocity function, both in units of $h^3$ Mpc$^{-3}$ dex$^{-1}$. 
To construct a mock galaxy catalog of luminosities and stellar masses 
from the BolshoiP simulation,
we apply the above procedure by using as input the global \ugriz\ GLFs and the \gsmf\ derived 
in Section \ref{global_density}.

Equation (\ref{AMT_Eq}) is the simplest form that SHAM could take as it ignores the existence of a physical scatter 
around the relationship between $\mathcal{P}_{\rm gal}$ and \vmax. Including physical scatter in Equation (\ref{AMT_Eq}) 
is no longer considered valid and should be modified accordingly \citep[for more details see][]{Behroozi+2010}. Constraints based on weak-lensing analysis
\citep{Leauthaud+2012}; satellite kinematics \citep{More+2009,More+2011}; and galaxy clustering \citep{Zheng+2007b,Zehavi+2011,Yang+2012} have shown that this is of the order of $\sim0.15$ dex in the
case of the stellar but similar in $r-$band magnitude. There are no constraints as for the dispersion around shorter wavelengths. In 
addition, it is not clear how to sample galaxy properties in a system with $n$ number of properties from 
the joint probability distribution ${\rm  prop}(\mathcal{P}_{{\rm gal},1}, ..., \mathcal{P}_{{\rm gal},n}| V_{\rm max}).$\footnote{In particular this paper uses
five bands $u$, $g$, $r$, $i$, and $z$ and a stellar mass \ms\ making a total of $n=6$.}  Instead of that, studies that aim at to constrain the galaxy-halo
connection use marginalization to constrain the probability distribution function  ${\rm  prop}(\mathcal{P}_{{\rm gal},i}| V_{\rm max})$ for $i$th galaxy property. 
In this paper we are interested in the statistical correlation of the galaxy-halo connection in which case Equation (\ref{AMT_Eq}) is
a good approximation. Studying and quantifying  the physical scatter around the relations is beyond the scope of this work.   
Also, ignoring the scatter around the galaxy-halo connection makes it easier to interpret. For those reasons we have opted to ignore the
any source of scatter in our relationships. 

%%%%%%%%%%%%%%%%%%%%%%%%%%%%%%%
\begin{figure*}
\includegraphics[height=3.8in,width=6.1in]{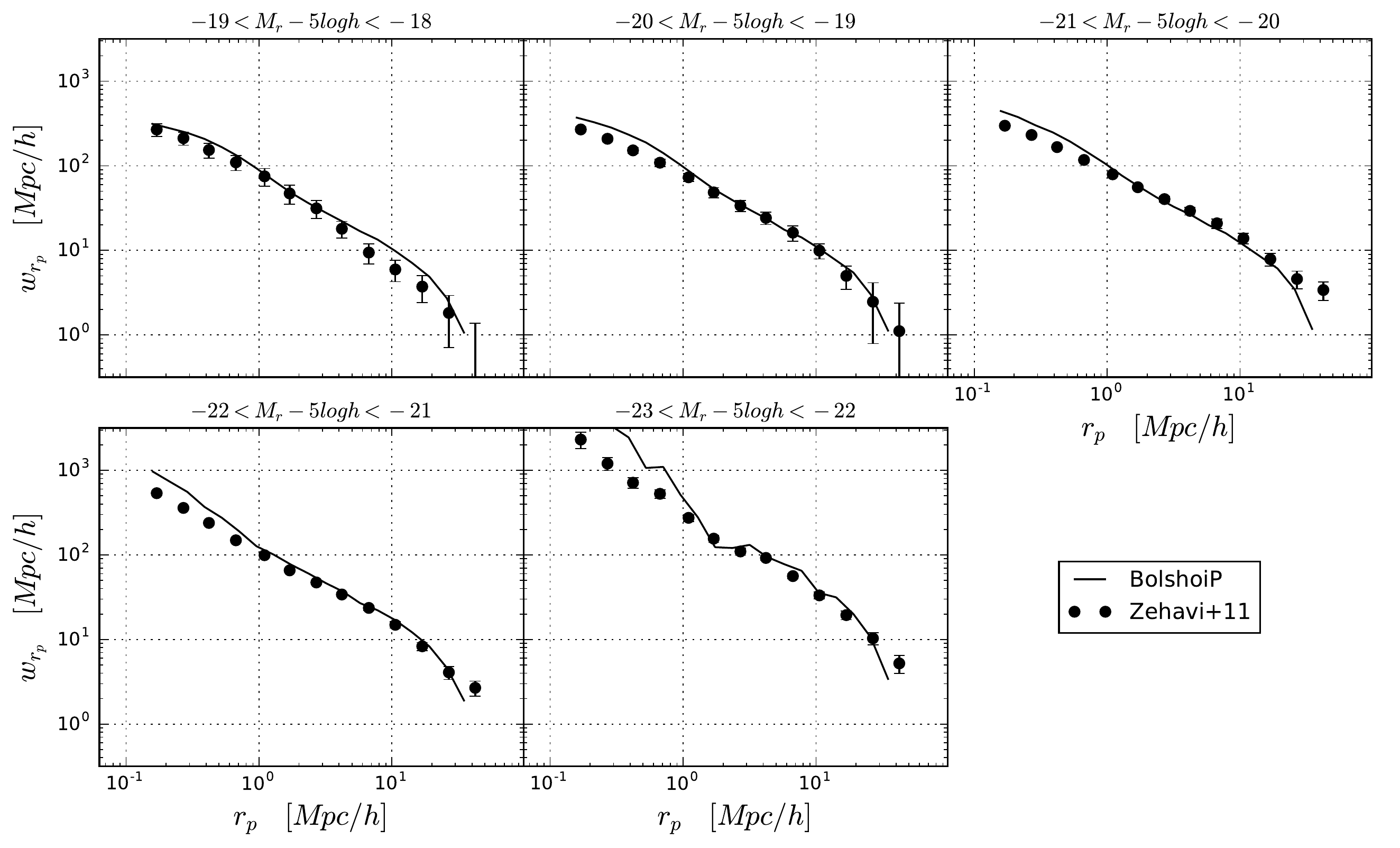}
\caption{Two-point correlation function in five luminosity bins at $z=0.1$. The solid lines show the predicted two-point correlation based on our $r$-band magnitude-to-\vmax\
relation from SHAM, while the circles with error bars show the same but for the SDSS DR7 \citep{Zehavi+2011}. 
 }
\label{wrp_lum}
\end{figure*}
%%%%%%%%%%%%%%%%%%%%%%%%%%%%%%%

%%%%%%%%%%%%%%%%%%%%%%%%%%%%%%
\begin{figure*}
\includegraphics[height=3.8in,width=6.1in]{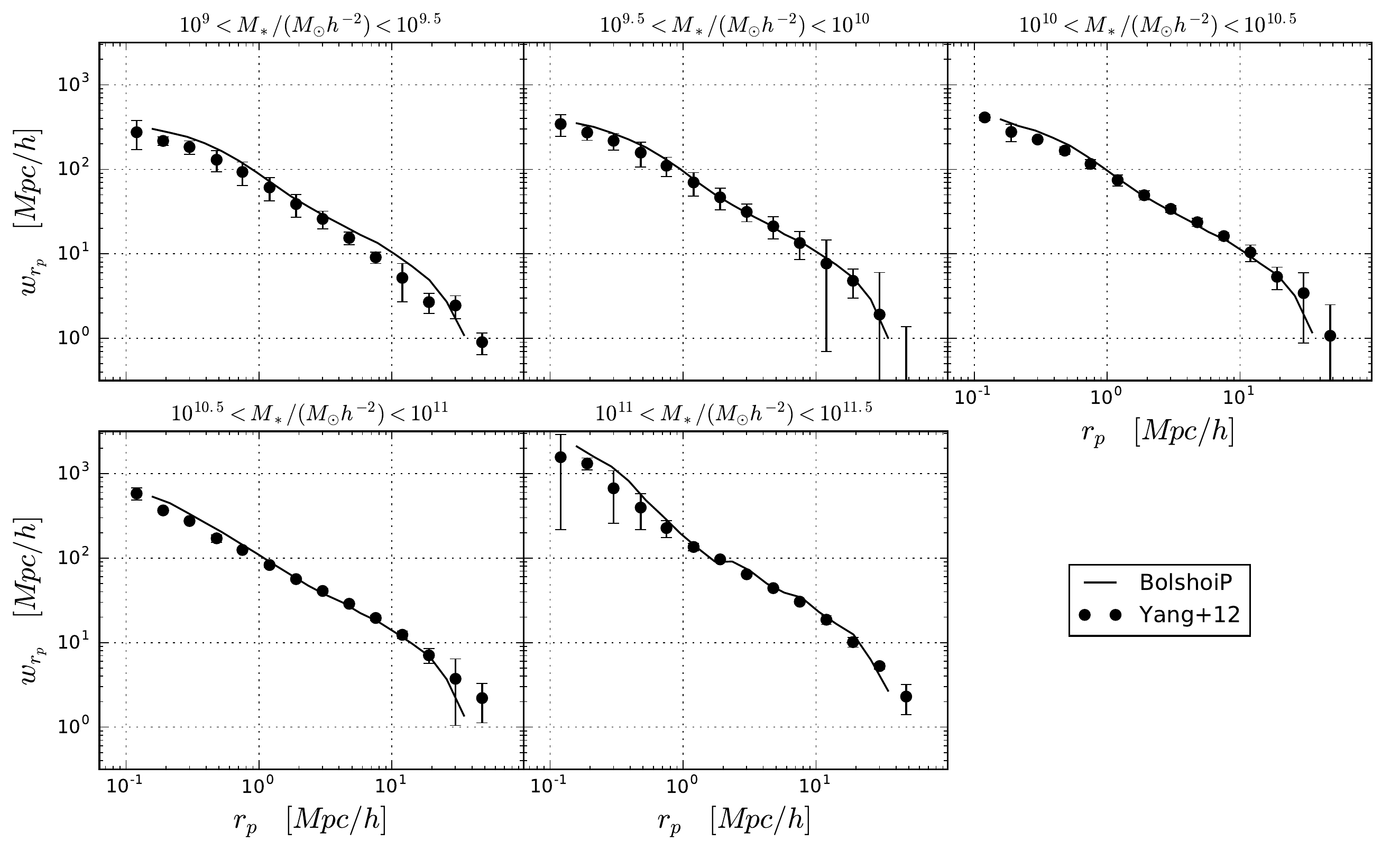}
\caption{Two-point correlation function in five stellar mass bins. The solid lines show the predicted two-point correlation based on our stellar mass-to-\vmax\
relation from SHAM, while the circles with error bars show the same but for SDSS DR7 \citep{Yang+2012}.
 }
\label{wrp_ms}
\end{figure*}
%%%%%%%%%%%%%%%%%%%%%%%%%%%%%%%

Previous studies have found that for distinct dark matter halos (those that 
are not contained in bigger halos), the maximum circular 
velocity \vmax\ is the halo property
that correlates best with the hosted galaxy's luminosity/stellar mass. 
This is likely because the properties of a halo's central region, where its central galaxy resides, are better described 
by $\vmax$ than $\mvir$.\footnote{For a NFW halo, $\vmax$ is reached at $R_{\rm max} = 2.16 R_s$, 
where $R_s$ is the NFW scale radius $R_s = R_{\rm vir}/C$ and $C$ is the NFW concentration \citep[e.g.,][]{Klypin+2001}. Since
$C \sim 10$ for Milky Way mass halos at $z=0$, $R_{\rm max} \sim (1/5) R_{\rm vir}$.}
%For subhalos (halos that are contained in bigger halos), however, 
By comparing to observations of galaxy clustering,
\citet{Reddick+2013} and more recently \citet{Campbell+2017}
have found that for subhalos, the property that correlates best with luminosity/stellar mass is
the highest maximum circular velocity 
reached along the main progenitor branch of the halo's merger tree. 
This presumably reflects the fact that subhalos can lose mass once they approach and fall into a larger halo, while the host galaxy at the halo's center is unaffected by this halo mass loss.
Thus, in this paper we use 
\begin{equation}
	\vmax = \left\{ 
			\begin{array}{c l}
				\vmax & \mbox{Distinct halos}\\
				\vpeak & \mbox{Subhalos}
			\end{array},\right.
\label{vmax-def}
\end{equation}
as the halo proxy for galaxy properties $\mathcal{P}_{\rm gal}$, where \vpeak\ is the maximum circular velocity throughout the entire 
history of a subhalo and \vmax\ is at the observed time for distinct halos. 

Figure \ref{galaxy_halo_connection} shows the relationships between galaxy luminosities
$u$, $g$, $r$, $i$, and $z$ and galaxy stellar masses to halo maximum circular velocities. Table \ref{galaxy_halo_table}, reports the values
from Figure \ref{galaxy_halo_connection}.
%and when using stellar mass as the galaxy property. 
Most of these relationships are steeply increasing
with \vmax\ for velocities below $\vmax\sim 160$ km s$^{-1}$. At higher velocities 
the relationships are shallower. The shapes of these relations are governed mostly by the
shapes of the GLFs and \gsmf, since the velocity function $\phi_V$ is approximately a power-law
over the range plotted in Figure \ref{galaxy_halo_connection}, see \citet{Rodriguez-Puebla+2016}.  

Note that at this point every halo and subhalo in the BolshoiP simulation at rest frame $z=0$
has been assigned a magnitude in the five bands $u$, $g$, $r$, $i$, and $z$ and a stellar mass \ms.
%A few words are worth of mentioning here. 
%Given that every halo in the BolshoiP has five magnitude bands, \ugriz, 
Therefore, one might be tempted to correlate galaxy colors such as red or blue (i.e. differences between galaxy magnitudes) with halo properties.
%in order to predict galaxies according their membership of being blue or red. 
If we did this, we would be ignoring the scatter around our luminosity/stellar mass-to-\vmax\ relationships, and
galaxies with the same magnitude or \ms\ would have the same color, contrary to observation.
Fortunately, 
%while we have checked that we are reproducing the average color-magnitude/stellar mass relation from the
%SDSS DR7, galaxies with the same \ms\ or magnitude will have the same color. In other words,
%there is no scatter in our color-magnitude/stellar mass relation. The main reason is that we had
%ignored the scatter around our lumonosity/stellar mass-to-\vmax\ relationships. 
including a scatter around those relationships will not impact our conclusions given that 
{\it i}) the scatter does not substantially impact the results presented in Figure \ref{galaxy_halo_connection} and 
{\it ii}) we are here interested only in the statistical correlation of the galaxy properties with environment. Nevertheless,
in Section \ref{satellites_predictions} we will study the statistical correlation between color and
environment for all galaxies, and separately for central and satellite galaxies. 

As a sanity check, we show that our mock galaxy catalog in the BolshoiP 
reproduces the projected two-point correlation function of SDSS galaxies.\footnote{When computing the  
projected two-point correlation function in the 
BolshoiP simulation, we integrate over the line-of-sight from $r_\pi = 0$ to $r_\pi = 40$ $h^{-1}$ Mpc, similarly 
to observations.} Figures \ref{wrp_lum} and \ref{wrp_ms} show, respectively, that this is the case for 
the $r$-band and stellar mass projected 
two point correlation functions. In the case of $r$-band, we compared to 
\citet{Zehavi+2011} who used $r$-band
magnitudes at $z=0.1$. We transformed our $r$-band magnitudes to $z=0.1$ by
finding the correlation between model magnitudes at $z=0$ and at $z=0.1$ from the tables 
of the \nyu\footnote{Specifically, we found that $M_r(z=0.1) = 0.992 \times M_r(z=0)+0.041$ with a Pearson correlation coefficient of $r=0.998$.}. 
For the projected two point correlation function in stellar mass bins we compare with 
\citet{Yang+2012}.

\subsection{Measurements of the mock \ugriz\ GLFs and the \gsmf\ as a function of environment}

%%%%%%%%%%%%%%%%%%%%%%%%%%%%%%%
\begin{figure*}
\includegraphics[height=3.8in,width=5.1in]{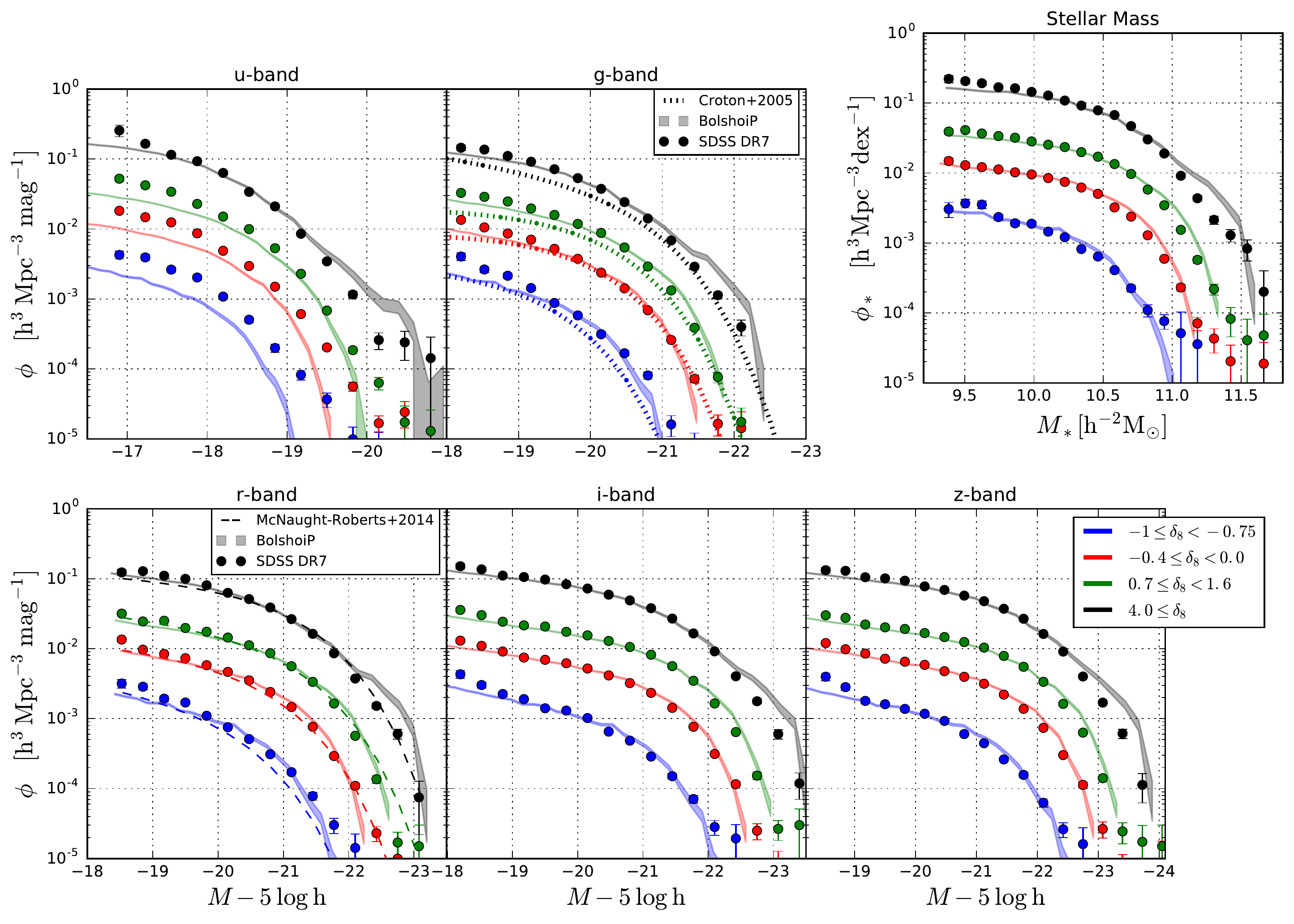}
\caption{Comparison between the observed SDSS DR7 \ugriz\ GLFs and \gsmf, filled circles with error bars, and the ones predicted based on the BolshoiP 
simulation from SHAM, shaded regions, at four environmental densities in spheres of radius 8 $h^{-1}$Mpc. We also reproduce the best fitting Schechter functions to the $r$-band GLFs from the GAMA survey \citep{McNaught-Roberts+2014}. Observe that
SHAM predictions are in excellent agreement with observations, especially for the longest wavelength bands and stellar mass. 
 }
\label{environment_GLF_SHAM}
\end{figure*}
%%%%%%%%%%%%%%%%%%%%%%%%%%%%%%%

%%%%%%%%%%%%%%%%%%%%%%%%%%%%%%%
\begin{figure*}
\includegraphics[height=2.8in,width=3.2in]{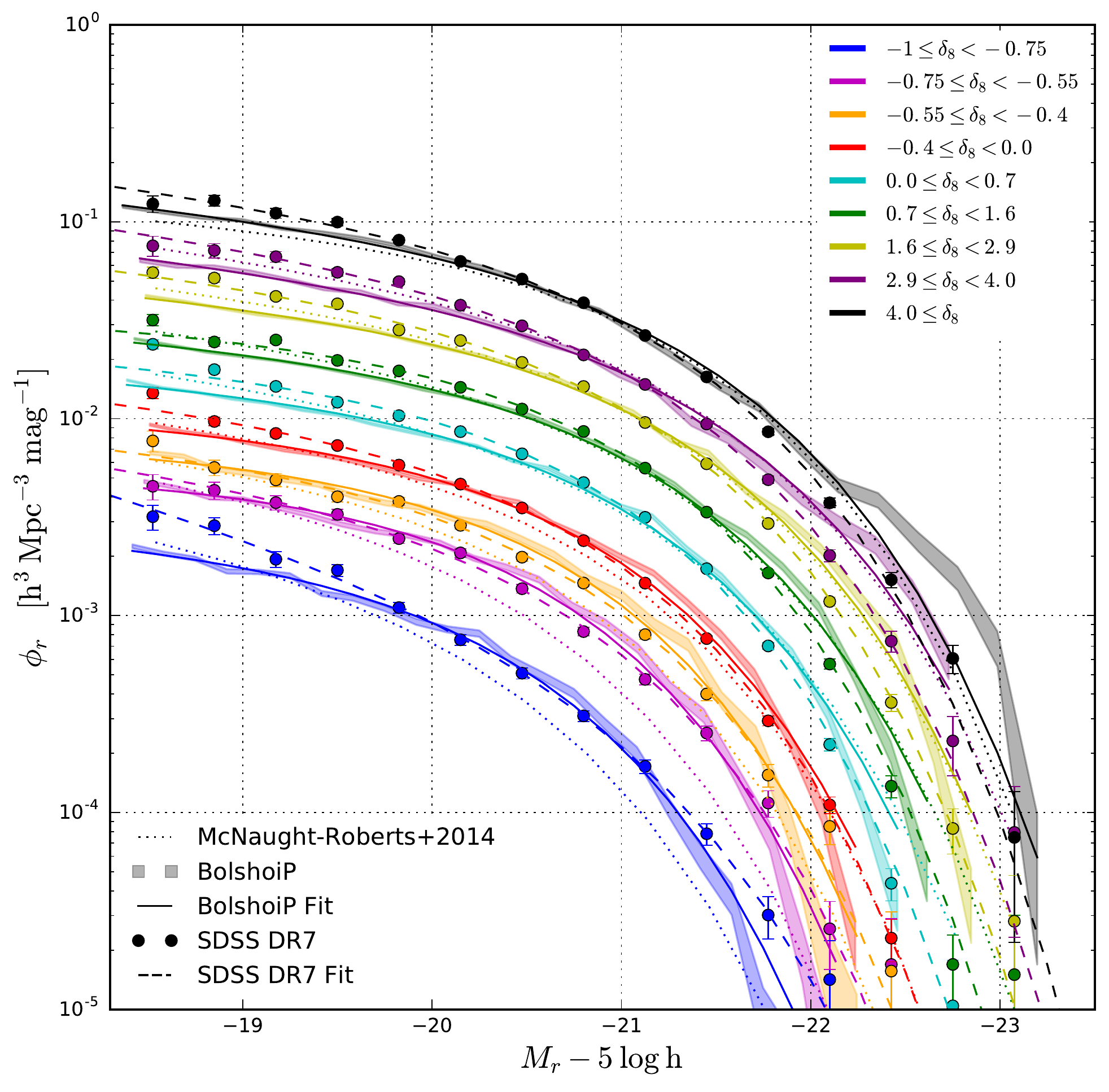}
\hspace*{10pt}
\includegraphics[height=2.8in,width=3.2in]{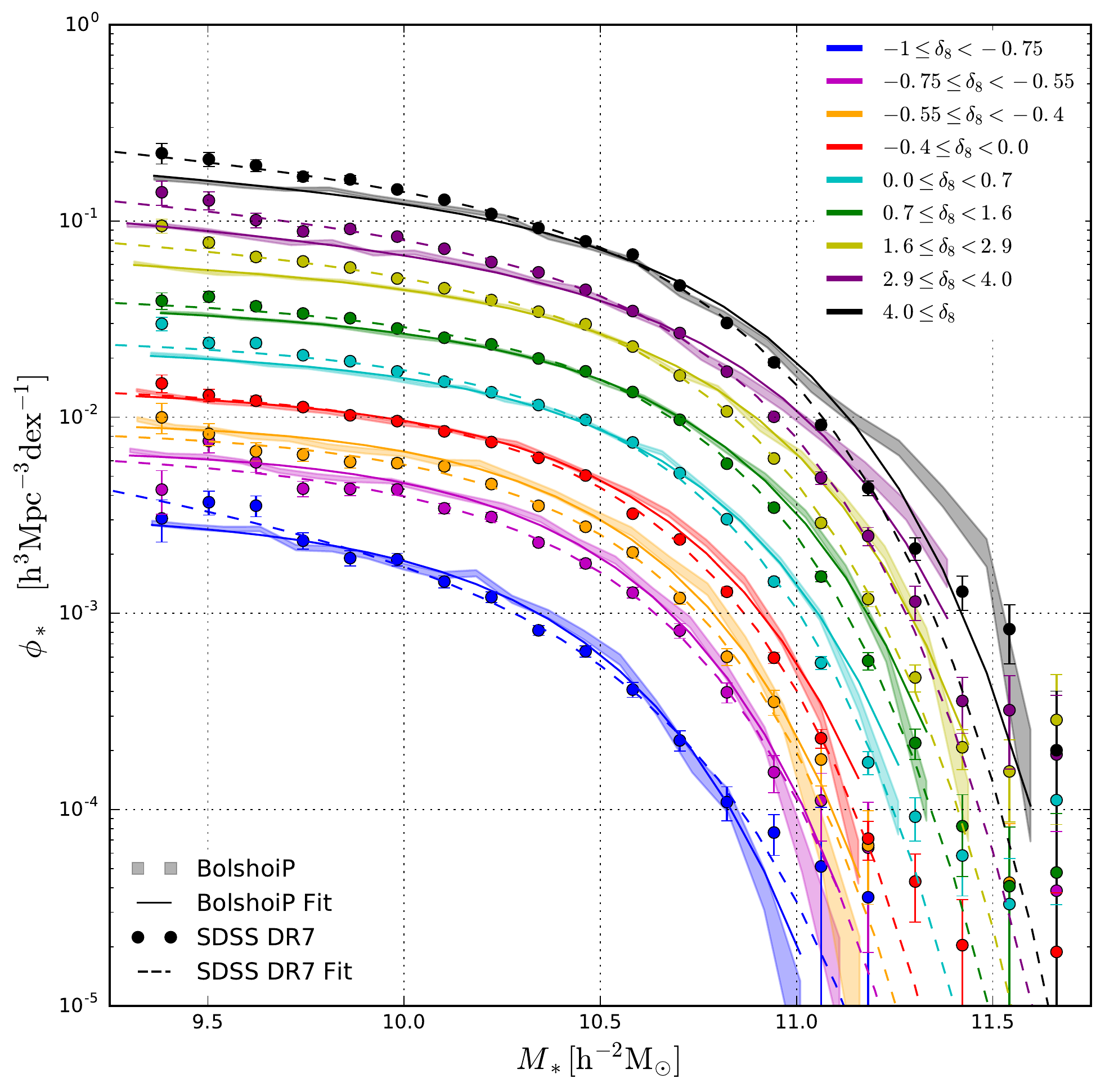}
\caption{{\bf Left Panel:} Comparison between the observed $r-$band GLF with environmental density in spheres of 8 $h^{-1}$Mpc, filled circles with error bars, 
and the ones predicted based on the BolshoiP simulation from SHAM, shaded regions. The dashed and solid lines show the best fitting Schechter 
functions to the observed and the mocked $r-$band GLFs while the dotted lines show the same but from the GAMA survey \citep{McNaught-Roberts+2014}.
{\bf Right Panel:} Similar to the left panel but for the \gsmf\ with environmental density. Here again the dashed and solid lines are the best fitting Schechter functions to
the observed and mocked GSMFs.}
\label{environment_GLF_rband_sham}
\end{figure*}
%%%%%%%%%%%%%%%%%%%%%%%%%%%%%%%

Our mock galaxy catalog is a volume complete sample down to halos of maximum circular velocity
$\vmax\sim55$ km$s^{-1}$, corresponding to galaxies brighter than $M_r-5\log h\sim-14$, see 
Figure \ref{galaxy_halo_connection}\footnote{In fact, the minimum halo allowed by the observations is for
halos above $\vmax\sim90$ km$s^{-1}$, corresponding to galaxies brighter than $M_r-5\log h\sim-17$, below this limit our mock catalog should be considered as
an extrapolation to observations.}. This magnitude
completeness is well above the completeness of the SDSS DR7. Thus, galaxies selected in the
absolute magnitude range $-21.8<M_r-5\log h<-20.1$ define a volume-limited DDP sample.
In other words, incompleteness is not a problem for our mock galaxy catalog. 
Overdensity and density contrast measurements for each mock galaxy in the BolshoiP simulation are obtained as 
described in Section \ref{obs_Over}.   

We estimate the dependence of the \ugriz\ GLFs with environment in our mock galaxy catalog as 
\begin{equation}
\phi_X(M_X|\delta_8) = \sum_{i=1}^N  \frac{\omega_i(M_X\pm \Delta M_X / 2|\delta_{r_8} \pm \Delta \delta_{r_8}/2)}{\Delta M_X  f_{\rm BP}(\delta_8)  L_{\rm BP}^3}.
\end{equation}
Here, $\omega_i = 1$ if a galaxy is within the interval $M_X\pm \Delta M_X / 2$ given that its overdensity is in the range 
$\delta_{r_8} \pm \Delta \delta_{r_8}/2$, otherwise it is 0. Again,  
$M_X$ refers to $M_u$, $ M_g$, $ M_r$, $ M_i$, $ M_z$ and $\log\ms$.
The function $f_{\rm BP}(\delta_8)$
is the fraction of effective volume by a given overdensity bin for the BolshoiP simulation. In order to
determine $f_{\rm BP}(\delta_8)$, we create a random catalog of $N_r\sim1.2\times10^{6}$ points in a box of side length identical to the BolshoiP simulation, i.e., 
$L_{\rm BP} = $ 250 $h^{-1}$Mpc. Using Equation (\ref{over_fraction8}) allows us to calculate $f_{\rm BP}(\delta_8)$.

\section{Results on Environmental Density Dependence}
\label{results}

%%%%%%%%%%%%%%%%%%%%%%%%%%%%%%%
\begin{figure*}
\hspace*{10pt}
\includegraphics[height=6in,width=3.in]{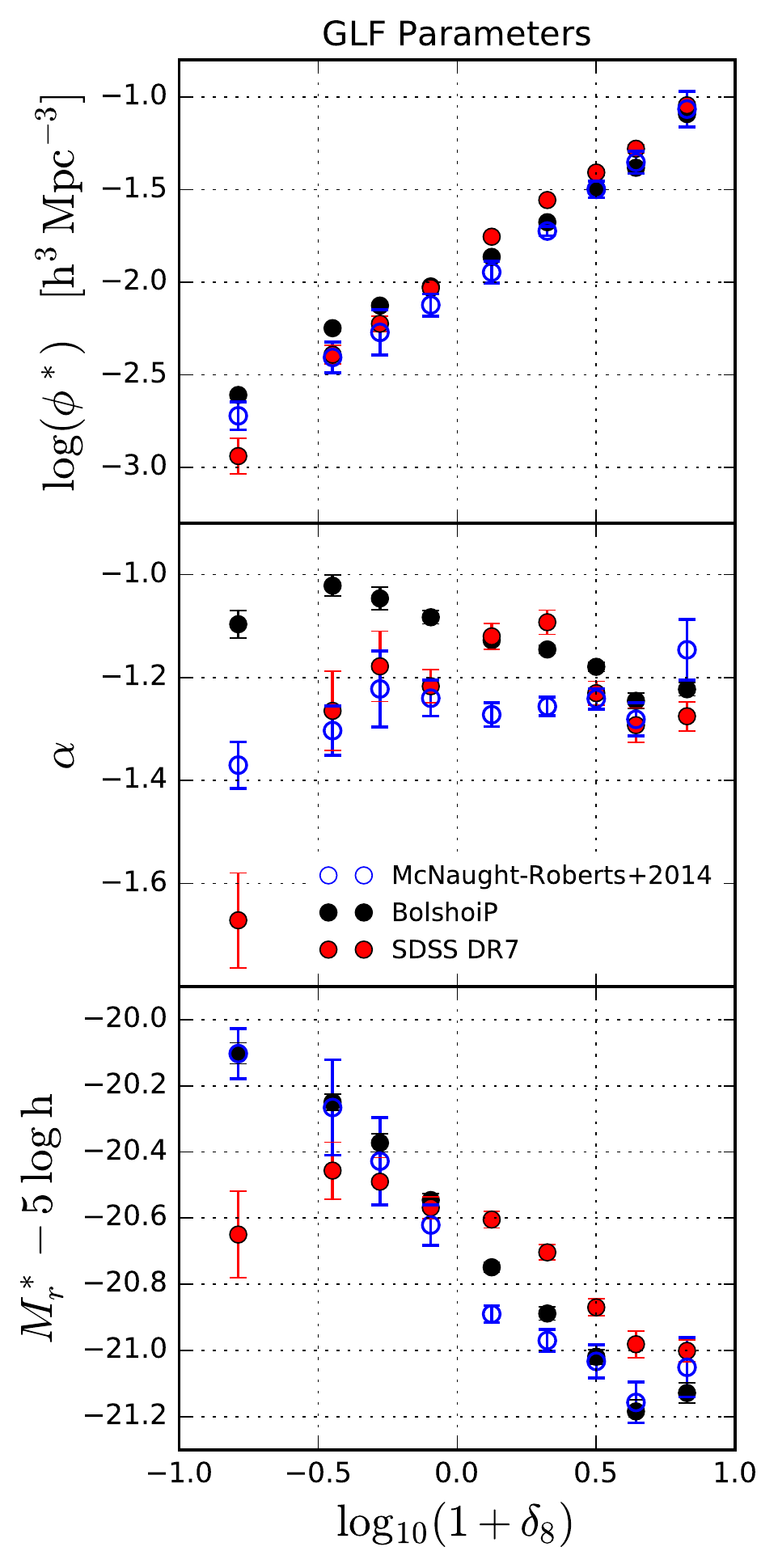}
\includegraphics[height=6in,width=3.in]{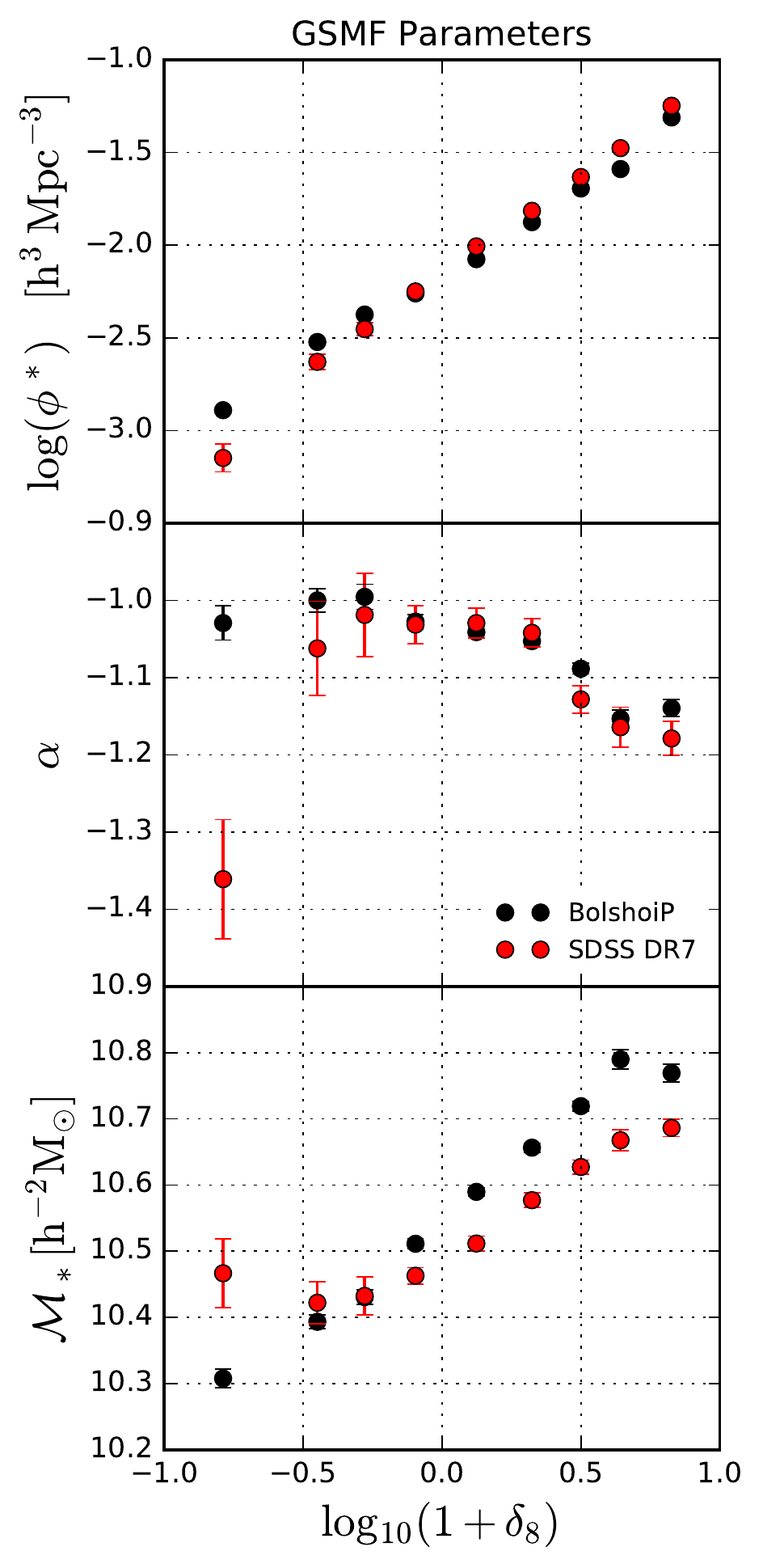}
%\vspace*{-100pt}
\caption{{\bf Left Panel:} The dependence of the $r-$band Schechter function parameters on environmental overdensity $\delta_8$ in spheres of 8 $h^{-1}$ Mpc  (Eq (\ref{delta_8})). {\bf Right Panel:}
The dependence of the galaxy stellar mass function Schechter parameters on environmental density. 
 }
\label{Boshoi_density_r_Sche_param}
\end{figure*}
%%%%%%%%%%%%%%%%%%%%%%%%%%%%%%%

In this section we present our determinations for the environmental density dependence of the \ugriz\ GLFs and the \gsmf\
from the SDSS DR7 and the BolshoiP. Here, we will investigate how well the assumption that the
statistical properties of galaxies are 
fully determined by $\vmax$ can predict  the dependence of the \ugriz\ GLFs and \gsmf\ with environment. 
We will show that predictions from SHAM are in remarkable agreement with 
the data from the SDSS DR7, especially for the longer wavelength 
bands. Finally, we show that SHAM also reproduces the correct 
dependence on environmental density of both the $r$-band GLFs and 
\gsmf\ for centrals and satellites, although 
it fails to reproduce the observed relationship between environment 
and color. 

\subsection{SDSS DR7}

Figure \ref{environment_GLF_SHAM} shows the dependence of the SDSS DR7 \ugriz\ GLFs as well as the \gsmf\ with environmental density measured in spheres of radius 8 $h^{-1}$Mpc. 
For the sake of the simplicity, we present only four overdensity bins in Figure \ref{environment_GLF_SHAM}.  %although 
In Figure \ref{environment_GLF_rband_sham} we show the determinations in nine density bins for the $r$-band GLFs and \gsmf. In order to compare with recent observational results we use 
identical environment density bins as in  \citet{McNaught-Roberts+2014},  who used galaxies from the GAMA survey to measure the dependence
of the $r$-band GLF on environment over the redshift range $0.04<z<0.26$ in spheres of radius of 8 $h^{-1}$Mpc. 

The  $r-$band panel of Figure \ref{environment_GLF_SHAM}  shows 
that our determinations are in good agreement with results from the GAMA survey.
% from the $r-$band magnitude. 
In the $g$-band panel of the same Figure, we present a comparison with the previously published results by \citet{Croton+2005}, who used 
the 2dF Galaxy Redshift Survey to measure the dependence of the $b_{\rm J}$-band GLFs at a zero redshift rest-frame in spheres of radius of 8 $h^{-1}$Mpc. We convert the $b_{\rm J}$-band GLFs from \citet{Croton+2005} to the $g$-band 
by applying a shift of -0.25 to their magnitudes, that is, $M_g = M_{b_{\rm J}} - 0.25$ \citep{Blanton+2005}. We observe
good agreement with the result of \citet{Croton+2005} for most of our density bins.  A better comparison would have used identical density bins; however, the density bins used by 
% for a more meaningful comparison. 
\citet{Croton+2005} are close to ours. 
Finally, in Figure \ref{environment_GLF_SHAM} we also extend previous results by 
presenting the GLFs for the $u$, $i$ and $z$ bands and for the \gsmf. 
We are not aware of any published low redshift GLFs for the $u$, $i$ and $z$ bands. 

The left panel of Figure \ref{environment_GLF_rband_sham} shows again the dependence of the 
GLF in the $r-$band, but now for all our overdensity bins, filled circles
with error bars. In order to report an analytical
model for the luminosity functions, we fit observations to a simple Schechter function; observations show that this model is a good description
for the data, given by 
	\begin{equation}
		\phi(M)=\frac{\ln 10}{2.5}\phi^*10^{0.4(M^*-M)\left(1+\alpha_1\right)}\exp\left(-10^{0.4\left(M^*-M\right)}\right),
	\end{equation} 
in units of $h^3$ Mpc$^{-3}$ mag$^{-1}$. 
The best fit to simple Schechter functions are shown as the dashed lines in the same plot, and we report 
the Schechter parameters as a function of the density contrast in the left panel of Figure \ref{Boshoi_density_r_Sche_param}. 
The best fitting parameters are listed
in Table \ref{T2}. For comparison, we reproduce the best fit to a Schechter function from 
 \citet{McNaught-Roberts+2014}, dotted lines.

Figure \ref{Boshoi_density_r_Sche_param} shows that
the normalization parameter of the Schechter function, $\phi^*$, depends strongly on density.
There are almost two orders of magnitude  
difference between the least and the highest density bins; see also Table \ref{T2}. In contrast, the faint-end slope, 
$\alpha$, remains practically constant with environment with a value of $\alpha = -1.0$ to $-1.2$. Note, however, that our analysis of the SDSS observations
shows that the GLF becomes steeper in the least dense environment, with $\alpha\sim-1.7$. The characteristic magnitude of the Schechter 
function, $M^*$, evolves only little with environment between $-1\lesssim\delta_8\lesssim0$ but it
increases above $\delta_8\sim0$. 
Finally, in the same figure, we reproduce the best fitting model parameters 
reported in \citet{McNaught-Roberts+2014}. In general, our determinations are in good agreement with 
the trends reported in \citet{McNaught-Roberts+2014} even at faint magnitudes, 
as is shown in Figures \ref{environment_GLF_rband_sham} and  \ref{Boshoi_density_r_Sche_param}.
This is reassuring since the GAMA survey is deeper than the SDSS which could result in a much better determination
of the faint-end. In addition, the subtended area
by the GAMA survey is much smaller than that of the SDSS, which could have resulted in GAMA underestimating  
%an impact in 
the abundance of massive galaxies 
%and also in themeasurements of their environments. 
in low-density environments. The reason for this is because the limited volume of GAMA does not adequately sample these rather rare galaxies 
in low-density regions.

%%%%%%%%%%%%%%%%%%%%%%%%%%%%%%%
\begin{figure*}
\includegraphics[height=5.5in,width=6in]{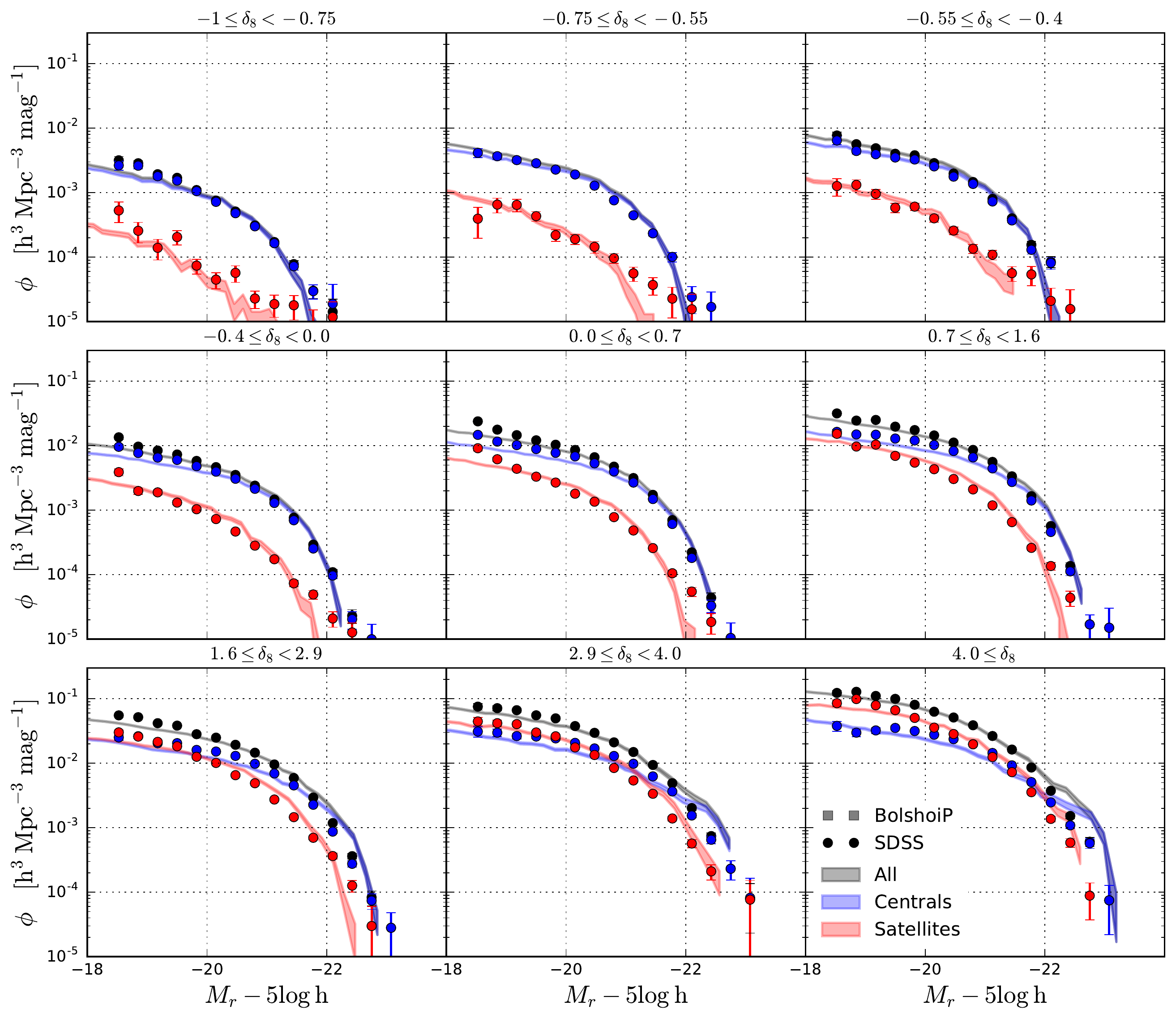}
\caption{The dependence on $r$-band magnitude of the GLFs in nine bins of environmental density in $8 h^{-1}$ Mpc spheres for all galaxies, central galaxies, and 
satellite galaxies. Filled circles with error bars show the results from the SDS DR7 while shaded areas
show the SHAM predictions from the BolshoiP simulation. There is a remarkable agreement 
between observations and SHAM predictions, even when dividing between centrals
and satellites. 
 }
\label{GLF_cs}
\end{figure*}
%%%%%%%%%%%%%%%%%%%%%%%%%%%%%%%

%%%%%%%%%%%%%%%%%%%%%%%%%%%%%%%
\begin{figure*}
\includegraphics[height=5.5in,width=6in]{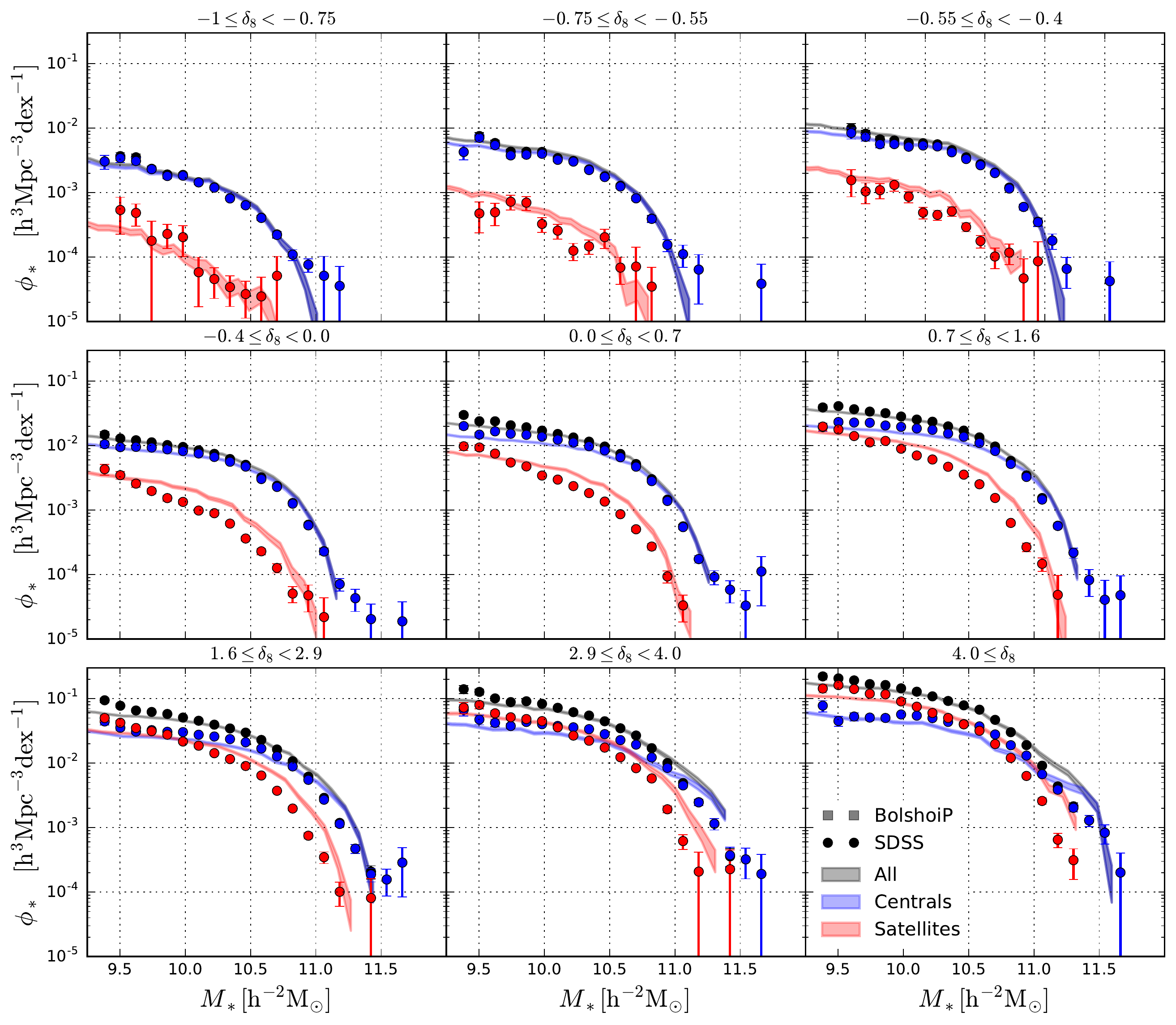}
\caption{Similarly to Figure \ref{GLF_cs} but for the \gsmf\ for all galaxies, central galaxies, and 
satellite galaxies. Filled circles with error bars show the results from the SDS DR7 while shaded areas
show the SHAM predictions from the BolshoiP simulation. While there is good agreement 
between observations and SHAM predictions for all galaxies and centrals, there is some tension 
with the SHAM predictions of the satellite \gsmf. 
 }
\label{GSMF_cs}
\end{figure*}
%%%%%%%%%%%%%%%%%%%%%%%%%%%%%%%

%%%%%%%%%%%%%%%%%%%%%%%%%%%%%%%
\begin{figure}
\vspace*{10pt}
\includegraphics[height=2.8in,width=3.2in]{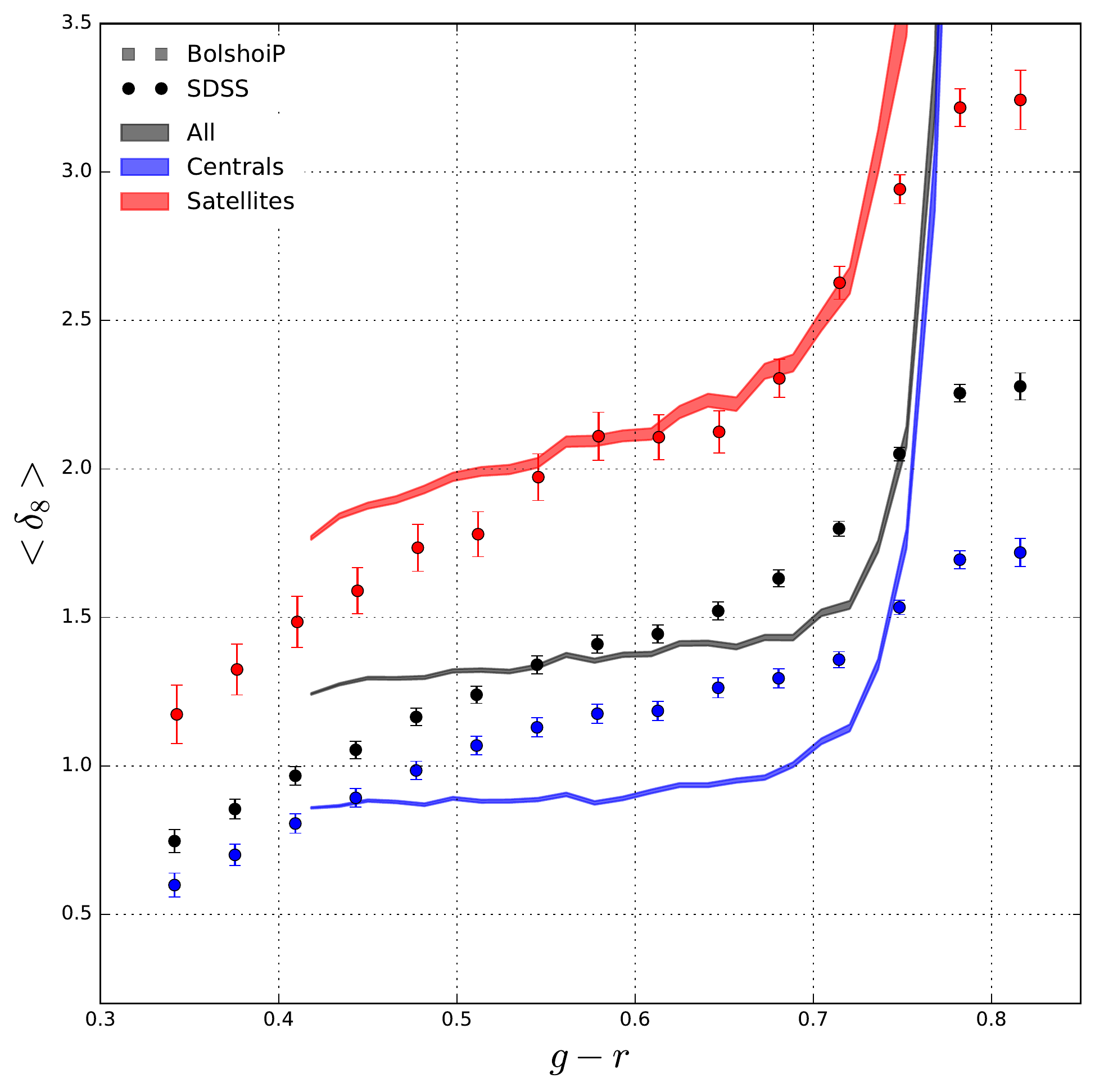}
\caption{Mean density in $8 h^{-1}$ Mpc spheres as a function of galaxy $g-r$ color, from the SDSS DR7 (shaded regions, representing the standard deviation) and the mean density predicted by SHAM based on the
BolshoiP simulation, filled circles with error bars. We present the mean density for all, central, and satellite galaxies as indicated
by the labels. SHAM fails to predict the correct relationship between mean density and galaxy colors
for all galaxies and central galaxies. In contrast, the SHAM prediction for satellite galaxies is in better agreement with observations. }
\label{color_environment}
\end{figure}
%%%%%%%%%%%%%%%%%%%%%%%%%%%%%%%

The right panel of Figure \ref{environment_GLF_rband_sham} shows the dependence of the $\gsmf$ for all our 
overdensity bins as well as their corresponding best fit to simple 
Schechter functions, filled circles with error bars and solid lines, respectively. In this case the Schechter function is given by
\begin{equation}
	\phi_*(\ms) =\phi_1^*\times{\ln 10}\times\left(\frac{\ms}{\mathcal{M}_1^*}\right)^{1+\alpha_1}\exp\left(-\frac{\ms}{\mathcal{M}_1^*}\right),
\end{equation}
with units of $h^3$ Mpc$^{-3}$ dex$^{-1}$. 
We report the best fitting parameters in Table \ref{T3}. 
The right panel of Figure \ref{Boshoi_density_r_Sche_param} presents the Schechter parameters for the $\gsmf$s as a function of the density contrast. Similarly to the 
GLFs, the normalization parameter for the \gsmf, $\phi^*$, depends strongly on density as a power-law and  
there are approximately two orders of magnitude difference between the GLFs in the least and the most dense environments. As for the faint-end slope, $\alpha$,
we observe that the general trend is that in high density environments the \gsmf\ becomes steeper than in low density environments. Nonetheless,
we observe, again, that in the lowest density bin the \gsmf\ becomes steeper than other density bins.
%contrary to the SHAM prediction. 
The characteristic stellar mass of the Schechter 
function, $\mathcal{M}_*$ 
increases with the environment at least for densities greater than $\delta_8\sim0$. In contrast, 
below $\delta_8\sim0$ it remains approximately constant.

%%%%%%%%%%%%TABLE%%%%%%%%%%%%%%%%%%%%%%
\begin{table*}
	\caption{Best fitting parameters from the observed SDSS DR7 GLFs}
	\begin{center}
		\begin{tabular}{c c c c c}
			\hline
						\multicolumn{5}{c}{Galaxy Luminosity Functions} \\  
			\hline
			\hline
						\multicolumn{5}{c}{SDSS DR7} \\  \\
			
			$\delta_{{\rm min},8}$ & $\delta_{{\rm max},8}$ & $\alpha$ & $M^*-5\log h$ & $\log\phi^*$ $\left[h^3 {\rm Mpc}^{-3} {\rm mag}^{-1}\right]$ \\
			\hline
			\hline
			-1 & -0.75 & $-1.671 \pm 0.092$ & $-20.650\pm0.131$ & $-2.938 \pm 0.095$ \\
			-0.75 & -0.55 & $-1.265 \pm 0.077$ & $-20.456\pm0.086$ & $-2.391 \pm 0.051$ \\
			-0.55 & -0.40 & $-1.178 \pm 0.068$ & $-20.490\pm0.073$ & $-2.224 \pm 0.041$ \\
			-0.40 & 0.00 & $-1.217 \pm 0.032$ & $-20.568\pm 0.033$ & $-2.031 \pm 0.019$ \\
			0.00 & 0.70 & $-1.120 \pm 0.025$ & $-20.604\pm 0.026$ & $-1.754 \pm 0.013$ \\
			0.70 & 1.60 & $-1.092 \pm 0.023$ & $-20.703\pm 0.025$ & $-1.556 \pm 0.012$ \\
			1.60 & 2.90 & $-1.230 \pm 0.033$ & $-20.870\pm 0.016$ & $-1.408 \pm 0.015$ \\
			2.90 & 4 & $-1.292 \pm 0.028$ & $-20.981\pm 0.040$ & $-1.280 \pm 0.023$ \\
			4.00 & $\infty$ & $-1.275 \pm 0.005$ & $-21.000\pm 0.034$ & $-1.045 \pm 0.020$ \\
			\hline
						\multicolumn{5}{c}{BolshoiP+SHAM} \\  \\
			\hline
			-1 & -0.75 & $-1.0963 \pm 0.0264$ & $-20.1018 \pm 0.0319$ & $-2.6084 \pm 0.0172$ \\
			-0.75 & -0.55 & $-1.0214 \pm 0.0206$ & $-20.2498 \pm 0.0238$ & $-2.2478 \pm 0.0127$ \\
			-0.55 & -0.40 & $-1.0462 \pm 0.0225$ & $-20.3727 \pm 0.0273$ & $-2.1260 \pm 0.0147$ \\
			-0.40 & 0.00 & $-1.0828 \pm 0.0129$ & $-20.5454 \pm 0.0193$ & $-2.0225 \pm 0.0096$ \\
			0.00 & 0.70 & $-1.1274 \pm 0.0087$ & $-20.7486 \pm 0.0142$ & $-1.8634 \pm 0.0075$ \\
			0.70 & 1.60 & $-1.1453 \pm 0.0086$ & $-20.8882 \pm 0.0204$ & $-1.6760 \pm 0.0079$ \\
			1.60 & 2.90 & $-1.1790 \pm 0.0092$ & $-21.0189 \pm 0.0220$ & $-1.4913 \pm 0.0090$ \\
			2.90 & 4 & $-1.2451 \pm 0.0154$ & $-21.1841 \pm 0.0354$ & $-1.3821 \pm 0.0179$ \\
			4.00 & $\infty$ & $-1.2227 \pm 0.0129$ & $-21.1285 \pm 0.0300$ & $-1.0924 \pm 0.0149$ \\
			\hline
		\end{tabular}
		\end{center}
	\label{T2}
\end{table*}
%%%%%%%%%%%%%%%%%%%%%%%%%%%%%%%%%%%%%

%%%%%%%%%%%%TABLE%%%%%%%%%%%%%%%%%%%%%%
\begin{table*}
	\caption{Best fitting parameters from the observed SDSS DR7 \gsmf}
	\begin{center}
		\begin{tabular}{c c c c c}
			\hline
						\multicolumn{5}{c}{Galaxy Stellar Mass Functions} \\  
			\hline
			\hline
						\multicolumn{5}{c}{SDSS DR7} \\  \\
			
			$\delta_{{\rm min},8}$ & $\delta_{{\rm max},8}$ & $\alpha$ & $\mathcal{M}_*$ $\left[h^{-2} \msun\right]$ & $\log\phi^*$ $\left[h^3 {\rm Mpc}^{-3} {\rm dex}^{-1}\right]$ \\
			\hline
			\hline
			-1 & -0.75 & $-1.361 \pm 0.077$ & $10.467\pm 0.052$ & $-3.148 \pm 0.075$ \\
			-0.75 & -0.55 & $-1.062  \pm 0.061$ & $10.422\pm 0.032$ & $-2.630 \pm 0.040$ \\
			-0.55 & -0.40 & $-1.019 \pm 0.054$ & $10.433\pm 0.029$ & $-2.453 \pm 0.035$ \\
			-0.40 & 0.00 & $-1.031 \pm 0.025$ & $10.463\pm 0.012$ & $-2.250 \pm 0.015$ \\
			0.00 & 0.70 & $-1.029 \pm 0.019$ & $10.511\pm 0.011$ & $-2.006 \pm 0.012$ \\
			0.70 & 1.60 & $-1.042 \pm 0.018$ & $10.577\pm 0.011$ & $-1.815 \pm 0.012$ \\
			1.60 & 2.90 & $-1.128 \pm 0.018$ & $10.627\pm 0.011$ & $-1.632 \pm 0.013$ \\
			2.90 & 4 & $-1.164 \pm 0.026$ & $10.668\pm 0.016$ & $-1.477 \pm 0.021$ \\
			4.00 & $\infty$ & $-1.179 \pm 0.022$ & $10.686\pm 0.013$ & $-1.248 \pm 0.017$ \\
			\hline
						\multicolumn{5}{c}{BolshoiP+SHAM} \\  \\
			\hline
			-1 & -0.75 & $-1.0293 \pm 0.0222$ & $10.3077 \pm 0.0142$ & $-2.8903 \pm 0.0176$ \\
			-0.75 & -0.55 & $-1.0000 \pm 0.0151$ & $10.3933 \pm 0.0105$ & $-2.5222 \pm 0.0128$ \\
			-0.55 & -0.40 & $-0.9954 \pm 0.0162$ & $10.4306 \pm 0.0110$ & $-2.3743 \pm 0.0138$ \\
			-0.40 & 0.00 & $-1.0275 \pm 0.0094$ & $10.5112 \pm 0.0074$ & $-2.2602 \pm 0.0088$ \\
			0.00 & 0.70 & $-1.0411 \pm 0.0074$ & $10.5896 \pm 0.0065$ & $-2.0754 \pm 0.0073$ \\
			0.70 & 1.60 & $-1.0528 \pm 0.0073$ & $10.6564 \pm 0.0069$ & $-1.8757 \pm 0.0075$ \\
			1.60 & 2.90 & $-1.0885 \pm 0.0069$ & $10.7191 \pm 0.0075$ & $-1.6939 \pm 0.0081$ \\
			2.90 & 4 & $-1.1533 \pm 0.0112$ & $10.7898 \pm 0.0146$ & $-1.5891 \pm 0.0154$ \\
			4.00 & $\infty$ & $-1.1396 \pm 0.0109$ & $10.7691 \pm 0.0135$ & $-1.3106 \pm 0.0143$ \\
			\hline
		\end{tabular}
		\end{center}
	\label{T3}
\end{table*}
%%%%%%%%%%%%%%%%%%%%%%%%%%%%%%%%%%%%%

\subsection{Comparison to Theoretical Determinations: SHAM}

Figure \ref{environment_GLF_SHAM} compares our observed \ugriz\ GLFs and 
the results derived from the mock galaxy sample based on SHAM.  
In general, we observe a remarkable agreement between observations and SHAM. This statement is true for all the 
luminosity bands as well as for stellar mass and for most of the density bins. 
This remarkable agreement is not a trivial result since we are assuming that \vmax\ fully determines
the magnitudes in the $u$, $g$, $r$, $i$, and $z$ bands and stellar mass \ms\ in every halo in the simulation. 
Additionally, in Section \ref{SHAM} we noted that the shape of the galaxy-halo connection is governed, mainly,
by the global shape of the galaxy number densities. Moreover, while we have defined our volume-limited DDP sample
as for the SDSS observations, it is subject to the assumptions behind SHAM as well. In addition, the real
correlation between $r$-band magnitude and all other galaxy properties is no doubt more complex than just monotonic relationships without scatters, as is derived in SHAM. 

Note, however, that there are some discrepancies towards bluer bands and low densities. %Ultimately,
Shorter wavelengths are more affected by recent star formation, and more likely to be related to halo mass accretion rates \citep[][and references therein]{SHARC}, while
infrared magnitudes depend more strongly on stellar mass. 
This perhaps 
just reflects that stellar mass is the galaxy property that most naturally correlates with \vmax. 
%rather than the limitations behind the assumptions in SHAM. 
Indeed, when comparing the 
environmental dependence of the observed and the mock \gsmf\ 
we observe, in general, rather good agreement. 

The left panel of Figure \ref{environment_GLF_rband_sham} compares the resulting dependence of the observed 
$r$-band GLFs with environment and the predictions based on SHAM for all our overdensity bins. 
This again shows the remarkable agreement between observations and SHAM for all our 
density bins.  Similarly, the right panel of Figure 
\ref{environment_GLF_rband_sham} compares the observed \gsmf\ with our predictions based on SHAM. 
Left panel of Figure \ref{Boshoi_density_r_Sche_param} compares the best fitting Schechter parameters 
for the $r$-band 
magnitude while the right panel shows the same but for stellar masses. 
In order to make a meaningful comparison with observations, we fit
the observed GLFs and \gsmf\ of the SDSS DR7 over the same dynamical ranges. In general,
we observe a good agreement between predictions from SHAM and the results from the SDSS DR7. 

While Figure \ref{environment_GLF_rband_sham} shows that the general trends are well predicted by SHAM, there are some differences 
that are worth discussing. SHAM is able to recover the overall normalization of the $r$-band GLF 
and the GSMF, but it slightly underpredicts the number of faint galaxies
and it also underpredicts the high-mass end in low-density environments. In high-density environments 
SHAM overpredicts the number of galaxies at the high mass end. A natural explanation for these 
differences could be the dependence of the galaxy-halo connection with environment.  
Recall that we are assuming zero scatter in the galaxy-halo connection. 
Despite the differences noted above, the extreme simplicity of SHAM captures extremely well
the dependences with environmental density of the galaxy distribution. This is remarkable and, as we noted earlier,
it might not be expected since halo properties depend on the local environment as well as the
large-scale environment. In order to understand better the success of SHAM and the nature 
of the above discrepancies, 
we now turn our attention to the dependence with environment of the $r$-band GLFs and GSMF of central and satellite galaxies separately. 

\subsubsection{SHAM Predictions for the Central and Satellite GLF and \gsmf}
\label{satellites_predictions}

Figures \ref{GLF_cs} and \ref{GSMF_cs} show respectively the dependence on environmental density of the 
$r$-band GLFs and \gsmf\ for all galaxies, and separately for centrals and satellites. The circles with error bars
show the results when using the memberships from the 
SDSS DR7 \citet{Yang+2012} galaxy group catalog, while the shaded areas
show the predictions from SHAM based on the BolshoiP simulation. 
When dividing the population between centrals and satellites, in general 
SHAM captures the observed dependences from the SDSS DR7 \citet{Yang+2012} galaxy group catalog. 
This simply reflects that the fraction of subhalos increases as a function of the environment as well
as the chances of finding high mass (sub)halos in dense environments. 

The agreement is particularly good for centrals. 
%than for satellite galaxies. 
However, the satellite $r$-band GLF is in much better agreement with observations 
than the satellite \gsmf: for
the $r$-band GLF we observe a marginal difference only, while SHAM predicts that there are more galaxies around the knee of the \gsmf\ compared to what is observed. It is not clear why we should expect this difference, but 
a potential explanation could be that satellite galaxies are much more sensitive to their local environment and to the definition of the DDP population. 
To help build intuition, recall that SHAM assigns every halo in the simulation five magnitudes in the $u$, $g$, $r$, $i$, and $z$ bands and stellar mass \ms.
Consider now that the relationship between $r$-band magnitude and all other galaxy properties are just monotonic 
and with zero scatter, as explained earlier. Thus, this oversimplification of much more complex relationships is
affecting the measurements of environmental dependences in satellite galaxies especially when these are projected in other observables.
Possibly, when using a stellar mass-based DDP population the 
problem will be inverted. In other words, we might observe a marginal difference between SHAM and the
$\gsmf$s but a larger differences between SHAM and the GLFs.  Of course
central galaxies are not exempt from also being affected, but given the good agreement with observations we conclude that the 
effect is only marginal. Another possible explanation is 
that the assumption of identical relations between centrals and satellites is more valid for the $r$-band luminosity than for
the stellar mass.  That is, the
stellar mass of satellite galaxies perhaps varies more strongly with \vmax\ than the $r$-band luminosity does. 

%{\bf 
A third possible explanation 
is that group finding algorithms are subject to errors. In a recent paper, \citet{Campbell+2015} showed that there are two main sources of errors
that could affect the comparison in Figures \ref{GLF_cs} and \ref{GSMF_cs}: {\it i)} central/satellite designation and {\it ii)} group membership determination.
In that paper, the authors showed that the \citet{Yang+2007} group finder
algorithm tends to misidentify central galaxies systematically with increasing group mass. In other words, satellites are sometimes mistakenly identified as centrals. Consequently, the GLFs and the \gsmf\ for centrals and satellites will be affected towards the bright-end. Note, however, that \citet{Campbell+2015} showed that for each satellite that is misidentified as a central, approximately a central is misidentified as a satellite in the \citet{Yang+2007} group finder. Thus, in the \citet{Yang+2007} group finder central/satellite designation is the main source of error rather than the group membership determination. While this a source of error that should be taken into account in our analysis, it is likely that this is not the main source of difference between observations and SHAM predictions. The reason is that there exists the above compensation effect in the identification of centrals and satellite galaxies which could leave, perhaps, the GLFs and the \gsmf\ of centrals and satellites with little or no changes.
%}

Finally, as we noted earlier, Figure \ref{environment_GLF_rband_sham} shows that SHAM overpredicts the number of high mass 
galaxies in high mass bins. Figures \ref{GLF_cs} and \ref{GSMF_cs} show that this excess of galaxies is due to central 
galaxies. We will discuss this in the light of the dependence of the galaxy-halo connection with environment in 
Section \ref{summ_discussion}. 

\subsubsection{The Relationship Between Color and Mean Environmental Density} 
 
Figure \ref{color_environment} shows the mean density as a function of the $g-r$ color
separately for all galaxies, centrals, and satellites. The filled circles with error bars show the mean density measured from 
the \citet{Yang+2012} galaxy group catalog while the shaded areas show the same but for the BolshoiP 
simulation. SHAM is unable to predict the correct correlation between mean density and galaxy colors
for all and central galaxies. SHAM predicts that, statistically speaking, 
the large-scale mean environmental density varies little with the colors of central galaxies, except that the reddest galaxies on average lie in the densest environments. 
Actually, this is not surprising since we assumed that the \ugriz\ bands and stellar mass are independent of environment when constructing our mock galaxy catalog and the above
is simply showing that one halo property does not fully determine the statistical properties of the galaxies. Other halo properties
that vary with environment should instead be used in order to reproduce the correct trends with environment.  
%{\bf 
Extensions to SHAM in which halo age is matched to galaxy age/color at a fixed luminosity/stellar mass and halo mass 
\citep[see e.g.,][]{HearinWatson2013,Masaki+2013} are promising approaches that could help to better explain the trends with observations.
%}
Nonetheless,  
SHAM predictions are in better agreement with the observed correlation of density with color for
satellite galaxies.

\section{Summary and Discussion}
\label{summ_discussion}

Subhalo abundance matching (SHAM) makes the assumption that
{\it one (sub)halo property fully determines the statistical
properties of their host galaxies}. Therefore, SHAM implies that {\it i}) the
galaxy-halo connection is identical between halos and subhalos, and
%{\it ii}) galaxy properties are independent of their local as well as their
%large-scale environment, no matter whether a galaxy is central or satellite. 
{\it ii}) the dependence of galaxy properties on environmental density comes 
entirely from the corresponding dependence on density of this (sub)halo property.
The halo property that this paper explores for SHAM is the quantity \vmax, which is defined in Equation (\ref{vmax-def}) 
as the maximum circular velocity for distinct halos, while for subhalos it the peak maximum circular velocity \vpeak\
reached along the halo's main progenitor branch. This is the most robust 
halo and subhalo property for SHAM \citep[see, e.g.,][]{Reddick+2013,Campbell+2017}. 
The galaxy properties we studied are the \ugriz\ GLFs 
as well as the \gsmf, which we determined from the SDSS DR7. We compared these observations with
SHAM predictions from a mock galaxy catalog based on the BolshoiP simulation \citep{Klypin+2016,Rodriguez-Puebla+2016}.
SHAM assigns every halo in the BolshoiP simulation magnitudes in the five SDSS
bands $u$, $g$, $r$, $i$ and $z$ and also a stellar mass \ms (Figure \ref{galaxy_halo_connection} and Appendix). 
We tested the assumptions behind SHAM by comparing the predicted and observed dependence of the 
\ugriz\ GLFs as well as the \gsmf\ on the environmental density from the SDSS DR7 \citet{Yang+2012} 
galaxy group catalog. The main results and conclusions are as follows:
\begin{itemize}
\item In general, the environmental dependence
of the \ugriz\ GLFs predicted by SHAM are in good 
agreement with the observed dependence from the SDSS DR7. This is especially
true for $r$ and infrared bands. Theoretically the stellar mass is the galaxy property
that is expected to depend more strongly on halo \vmax, while bluer bands also reflect recent effects
of star formation. 
\item We show that the environmental dependence of the \gsmf\ predicted 
by SHAM is in remarkable agreement with the observed dependence from the SDSS DR7,
reinforcing the above conclusion. 
\item When dividing the galaxy population into centrals and satellites SHAM predicts 
the correct dependence of the observed $r$-band GLF and \gsmf\ for centrals and 
satellite galaxies from the \citet{Yang+2012} group galaxy catalog. 
\item While SHAM predicts GLFs and the \gsmf\ that are in remarkable agreement 
with observations even when the galaxy population is separated between centrals
and satellites, SHAM does not predict the observed average relation between $g-r$ color and mean
environmental density. This is especially true for central galaxies, while the correlation
obtained for satellite galaxies is in better agreement with observations. 
\end{itemize}

Many previous authors have studied the correlation between galaxies and dark matter halos with environment both theoretically and observationally
\citep[see, e.g.,][and many more references cited therein]{Avila-Reese+2005,Baldry+2006,Blanton+2007,Maulbetsch+2007,Tinker+2011,Lacerna+2014,Lee+2017,Yang+2017}.
While most of these authors have focused on understanding this correlation by studying the 
galaxy distribution as a function of color, star formation or age and environment at a fixed \ms, here 
we take a different approach and exploit the extreme simplicity of SHAM. Firstly,
there are no special galaxies in SHAM. Second, SHAM can be applied to any galaxy property distribution. 
Thus, in our framework a halo and a subhalo with identical
\vmax\ will host galaxies with identical luminosities and stellar mass, no 
matter the halo's environmental density or position in the cosmic web. Our results are consistent with previous findings 
that halo \vmax\ could be enough to determine the luminosities and stellar masses. 
%Nevertheless, there exists the possibility that the dependence of the \ugriz\ GLFs and \gsmf\ with environment
%weakly constraint this assumption, so the above statement is not conclusive. 
%This could be the case as we have
However, we have also shown that SHAM is unable to reproduce the correct correlation between galaxy color and the mean density $\delta_8$ on a scale of 8 $h^{-1}$ Mpc. This result implies that additional halo properties that depend in some way on the
halo environment \citep[e.g.,][]{Lee+2017} should be employed to correctly reproduce 
the relationship between $\delta_8$ and
galaxy color.

Does the above imply that the galaxy-halo connection should depend on environment? 
On the one hand, from observations we have learned that 
the statistical properties of the galaxies such as color and star formation change with environment in the direction that low 
density environments are mostly populated by blue/star-forming galaxies while dense environments are mostly populated 
with red/quenched galaxies \citep[see for e.g.,][]{Hogg+2003,Baldry+2006,Tomczak+2017}. On the other hand,
the shape of the luminosity-to-\vmax\ and the stellar mass-to-\vmax\ relations, Figure \ref{galaxy_halo_connection}, 
contain information about the process that regulated the star formation in galaxies. Therefore, it is not a bad idea to consider that
the differences described in Figure \ref{environment_GLF_rband_sham} are the result that the galaxy-halo connection could
change with environment. 
For the sake of the simplicity, consider the \gsmf\ of central galaxies derived in the case of zero scatter around the 
$\ms=\ms(\vmax)$ relationship. Therefore, Equation (\ref{AMT_Eq}) can be rewritten to give the \gsmf\ as
\begin{equation}
\phi_* (\ms) = \phi_V (\vmax(\ms)) \times \alpha_{\rm gal},
\end{equation}
while the dependence with environment of the \gsmf\ of central galaxies is given by
\begin{equation}
\phi_* (\ms|\delta_8) = \phi_V (\vmax(\ms)|\delta_8) \times \alpha_{\rm gal},
\end{equation}
where $\alpha_{\rm gal} \equiv d\log \vmax(\ms) / d\log \ms$ is the logarithmic slope of the $\ms=\ms(\vmax)$ relationship
assumed to be independent of environment. Next, consider the simplest case in which $\phi_V (\vmax|\delta_8)$ is a double
power law such that $\phi_V (\vmax|\delta_8) \propto V_{\rm max}^{\beta(\delta_8)}$ for $V_{\rm max}\ll V_{\rm max}^*(\delta_8)$ and
 $\phi_V (\vmax|\delta_8) \propto V_{\rm max}^{\gamma(\delta_8)}$ for $V_{\rm max}\gg V_{\rm max}^*(\delta_8)$ where 
$V_{\rm max}^*(\delta_8)$ is a characteristic velocity and we have emphasized that the parameters $\beta, \gamma$ and $V_{\rm max}^*$
depend on the environment. In order to simplify even further the problem, consider that the $\ms=\ms(\vmax)$ relationship is a 
power law relation at low masses with logarithmic slope $\alpha_{\rm gal, low}$ while at high masses it is also a power law with
logarithmic slope of $\alpha_{\rm gal, high}$ . 
Based on the above, we can write the dependence with environment of the \gsmf\ of central galaxies in the limit cases
\begin{equation}
	\phi_* (\ms|\delta_8) \propto \left\{ 
			\begin{array}{@{}l@{}l@{}}
				\alpha_{\rm gal,low}\times M_*^{\beta(\delta_8)/\alpha_{\rm gal,low}} & \mbox{if } V_{\rm max}\ll V_{\rm max}^*(\delta_8)\\
				\alpha_{\rm gal,high}\times M_*^{\gamma(\delta_8)/\alpha_{\rm gal,high}} & \mbox{if } V_{\rm max}\gg V_{\rm max}^*(\delta_8)
			\end{array}.\right.
			\label{gsmf_approx}
\end{equation}
Thus, if $\alpha_{\rm gal,low}$ and $\alpha_{\rm gal,high}$ are independent of environment the resulting shape and dependence of $\phi_* (\ms|\delta_8)$
with environment can be simply understood as the dependence with environment of the slopes $\beta$ and $\gamma$ of the halo velocity function. 

By looking to the least (voids-like) and highest (cluster-like) density environments from Figure \ref{GSMF_cs}, upper left and bottom right panels respectively, we can use the
above model in order to understand how the galaxy-halo connection may depend on environment. The voids-like \gsmf\ from Figure \ref{GSMF_cs} shows that SHAM tends to underpredict the 
number density of central galaxies both at the low and high mass ends. In other words, the slopes predicted by SHAM at the low and high mass ends are, respectively, 
too shallow and steep compared to observations. Inverting this would require, based on Equation (\ref{gsmf_approx}), to make the slopes $\alpha_{\rm gal,low}$ and
$\alpha_{\rm gal,high}$ shallower and steeper, respectively, to what we derived from SHAM, see the right panel of Figure \ref{galaxy_halo_connection}. This implies that 
in low density environments at a fixed \vmax\ halos had been more efficient in forming stars\footnote{Note that we are assuming that the zero point of the $\ms=\ms(\vmax)$
remains the same in both case. This is not a bad approximation since SHAM is able to recover the normalization of the \gsmf\ in all environments.} both for at the low and
high mass end. In contrast, the high density \gsmf\ from Figure \ref{GSMF_cs} shows that SHAM tends to overpredict the 
number density of central galaxies at the high mass end. In this case, we invert the above trend by making the high mass end slope $\alpha_{\rm gal,high}$ more shallow 
compared to what is currently derived from SHAM. This implies that the star formation efficiency has been suppressed in high mass halos residing in high density environments
%{\bf 
with respect to the predictions of SHAM.
%} 

The above limiting cases show that the galaxy-halo connection is expected to change with environment in the direction that halos in low density environment should be more
efficient in transforming their gas into stars while in high density environments halos have become more passive. This is indeed consistent with the color/star formation
trends that have been observed in large galaxy surveys. Of course, our discussion is an oversimplification, and in order to model exactly how the galaxy-halo connection 
depends on environment we would need to use the dependence of the \gsmf\ with environment as an extra observational constraint for the galaxy-halo connection. 
%{\bf 
In a recent series of papers \citet{Tinker+2017a,Tinker+2017b} and \citet{Tinker+2017c} studied the galaxy-halo connection in the light of the 
relation between the star formation and environment at a fixed stellar and halo mass, obtaining similar conclusions to ours. 
That is, above-average galaxies with above average star formation rates and high halo accretion rates live in underdense environments, while the increase of the observed quenched fraction of galaxies
from low-to-high density environments is consistent with the fact that halo formation has an impact on quenching the star formation at high masses and densities. See also \citet{Lee+2017}
for similar conclusions.
%}

Finally, we expect that at high redshift the assumptions from SHAM are likely to be closer to reality. The reason is that as the Universe ages, the 
cosmic web becomes more mature and the dependence of halo properties with environment become also stronger. As we showed here, while there 
are some differences with observations of local galaxies, those are small despite the extreme simplicity of the SHAM assumptions. Therefore, we expect that the 
galaxy-halo connection should depend less on environment at high redshifts, when 
environmental process have not played a significant role.  

%It will be interesting to interpret the above in the light of other popular moles 
%such as the ``age matching" \citep{HearinWatson2013} and the iHOD approach 
%by \citet{Zu+2016}.

\section*{Acknowledgments} 
We thank Vladimir Avila-Reese, Peter Behroozi, Avishai Dekel, Sandra Faber, David Koo, Rachel Somerville, Risa Wechsler, and Chandrachani Ningombam for useful comments and discussions. AR-P thanks the UC-MEXUS-CONACYT program for support at UCSC.  JRP and CTL acknowledge support from grants HST-GO-12060.12-A and HST-AR-14578.001-A. The authors acknowledge the UC MEXUS-CONACYT Collaborative Research Grant CN-17-125. We thank the NASA Advanced Supercomputer program for allowing the Bolshoi-Planck simulation to be run on the Pleiades supercomputer at NASA Ames Research Center. We also thank the anonymous Referee for a useful report which helped improve the presentation of this paper. Part of the material of this paper was presented as the Bachelor of Science senior thesis of Radu Dragomir.
%ADD ADDITIONAL THANKS?

\bibliographystyle{mn2efix}
\bibliography{Bibliography}

\appendix

\section{Tables for the galaxy-halo connection}
This Section reports our Luminosity-to-\vmax\ and stellar mass-to-\vmax\ relations from SHAM in Table \ref{galaxy_halo_table}. For galaxies, we utilized the best fitting parameters for the global \ugriz\ galaxy luminosity functions and stellar mass function reported in Table \ref{T1}. In the case of distinct halos, \vmax\ refers to the halo maximum circular velocity, while for satellites $\vmax$ represents the highest maximum circular velocity
reached along the subhalo's main progenitor branch  $\vpeak$. Note that the validation limits for our Luminosity-to-\vmax\ and stellar mass-to-\vmax\ determinations are due to the range of the observed galaxy number density which corresponds to halos above $\vmax\sim90$ km/s even if the BolshoiP simulations is complete up to $\vmax\sim55$ km/s. Below this limit the mock catalog should be considered as an extrapolation to observations. 

%%%%%%%%%%%%TABLE%%%%%%%%%%%%%%%%%%%%%%
\begin{table*}
	\caption{Luminosity-to-\vmax\ relations and stellar mass-to-\vmax\ relation from SHAM.}
	\begin{center}
		\begin{tabular}{c c c c c c c}
			\hline			
			$\vmax$ [km/s] & $\log (\ms /h^{-2}\msun)$ & $M_u - 5\log h$  & $M_g - 5\log h$  & $M_r - 5\log h$  & $M_i - 5\log h$  & $M_z - 5\log h$\\
			\hline
			\hline
      80.0000 & 7.96057 & -14.5647 & -15.8541 & -16.2580 & -16.4285 & -16.5291\\
      88.5047 & 8.26452 & -15.3542 & -16.5009 & -16.9221 & -17.0909 & -17.2073\\
     97.9136 & 8.59933 & -15.9608 & -17.1241 & -17.5814 & -17.7615 & -17.9085\\
     108.323 & 8.95125 & -16.4419 & -17.6885 & -18.1921 & -18.3951 & -18.5809\\
     119.838 & 9.28607 & -16.8340 & -18.1794 & -18.7281 & -18.9579 & -19.1805\\
     132.578 & 9.57433 & -17.1613 & -18.5995 & -19.1867 & -19.4408 & -19.6931\\
     146.672 & 9.81092 & -17.4401 & -18.9593 & -19.5777 & -19.8514 & -20.1260\\
     162.265 & 10.0045 & -17.6817 & -19.2700 & -19.9132 & -20.2022 & -20.4935\\
     179.515 & 10.1654 & -17.8942 & -19.5413 & -20.2044 & -20.5053 & -20.8089\\
     198.599 & 10.3019 & -18.0835 & -19.7809 & -20.4603 & -20.7703 & -21.0832\\
     219.712 & 10.4197 & -18.2542 & -19.9950 & -20.6877 & -21.0049 & -21.3250\\
      243.070 & 10.5232 & -18.4096 & -20.1882 & -20.8920 & -21.2150 & -21.5406\\
      268.910 & 10.6154 & -18.5525 & -20.3641 & -21.0774 & -21.4049 & -21.7350\\
     297.497 & 10.6985 & -18.6852 & -20.5256 & -21.2471 & -21.5784 & -21.9119\\
     329.124 & 10.7742 & -18.8093 & -20.6752 & -21.4038 & -21.7381 & -22.0745\\
     364.113 & 10.8440 & -18.9266 & -20.8147 & -21.5495 & -21.8864 & -22.2252\\
     402.821 & 10.9088 & -19.0385 & -20.9457 & -21.6861 & -22.0252 & -22.3658\\
     445.645 & 10.9696 & -19.1464 & -21.0697 & -21.8151 & -22.1560 & -22.4983\\
     493.021 & 11.0270 & -19.2516 & -21.1879 & -21.9378 & -22.2803 & -22.6238\\
     545.433 & 11.0818 & -19.3557 & -21.3013 & -22.0554 & -22.3992 & -22.7438\\
     603.418 & 11.1344 & -19.4603 & -21.4109 & -22.1688 & -22.5139 & -22.8594\\
     667.566 & 11.1854 & -19.5678 & -21.5177 & -22.2792 & -22.6253 & -22.9716\\
     738.534 & 11.2352 & -19.6810 & -21.6226 & -22.3875 & -22.7345 & -23.0813\\
     817.047 & 11.2843 & -19.8042 & -21.7263 & -22.4944 & -22.8422 & -23.1896\\
     903.907 & 11.3331 & -19.9425 & -21.8298 & -22.6009 & -22.9495 & -23.2972\\
        1000.00 & 11.3820 & -20.1006 & -21.9337 & -22.7078 & -23.0570 & -23.4050\\
     1106.31 & 11.4313 & -20.2772 & -22.0389 & -22.8158 & -23.1656 & -23.5138\\
     1223.92 & 11.4815 & -20.4630 & -22.1461 & -22.9259 & -23.2762 & -23.6245\\
     1354.03 & 11.5328 & -20.6494 & -22.2561 & -23.0386 & -23.3894 & -23.7377\\
     1497.98 & 11.5856 & -20.8337 & -22.3695 & -23.1548 & -23.5060 & -23.8542\\
     1657.23 & 11.6403 & -21.0165 & -22.4871 & -23.2752 & -23.6267 & -23.9746\\
			\hline
		\end{tabular}
		\end{center}
	\label{galaxy_halo_table}
\end{table*}

\label{lastpage}
\end{document}